\documentclass[aps,prb,twocolumn,superscriptaddress,floatfix,amsmath,onlytext,
fullfamily,textlf,longbibliography]{revtex4-2} 

\usepackage[normalem]{ulem}
\usepackage{bm} 
\usepackage{amssymb}
\usepackage{amsfonts} 
\usepackage{amsmath} 

\usepackage{graphicx} 

\usepackage{xcolor} 
 
\usepackage[pdftex,colorlinks=true,allcolors=blue,breaklinks=true]{hyperref}  

\usepackage[utf8]{inputenc} 
\usepackage{textgreek}
 
\definecolor{g-blue}{rgb}{0.83,0.95,1}
\definecolor{g-yellow}{rgb}{1,1,0.7}
\definecolor{g-green}{rgb}{0.9,1,0.9}
\definecolor{green}{rgb}{0,0.6,0}
\definecolor{cyan}{rgb}{0,0.7,0.7}
\definecolor{black}{rgb}{0,0,0}
\definecolor{grey}{rgb}{0.4,0.4,0.4}
\definecolor{nature-blue}{rgb}{0.0,0.200,0.500}

\def \ed {\end{document}}
\def\Fbox#1{\vskip1ex\hbox to 8.5cm{\hfil\fboxsep0.3cm\fbox{%
		\parbox{8.0cm}{#1}}\hfil}\vskip1ex\noindent}  

\def\be{\begin{equation}}
\def\ee{\end{equation}}
\def\bea{\begin{eqnarray}}
\def\eea{\end{eqnarray}}
\def\bse{\begin{subequations}}
\def\ese{\end{subequations}}

\let \= \equiv  
\let\*\cdot 
\let\^\widehat 
\let\-\overline

\def\1{\bm1}

\def\<{\left\langle}    \def\>{\right\rangle}
\def\[ {\left[}         \def\]{\right]}

\begin{document} 
\title{Unified   
theory of  magnetization temperature dependence in  ferrimagnets
} 
    \author{Rostyslav O. Serha}	
    \email{rostyslav.serha@univie.ac.at}	
    \affiliation{Faculty of Physics, University of Vienna, 1090 Vienna, Austria}
    \affiliation{Vienna Doctoral School in Physics, University of Vienna, 1090 Vienna, Austria}
    \author{Anna Pomyalov}
    \email{anna.pomyalov@weizmann.ac.il}
    \affiliation{Department of Chemical and Biological Physics, Weizmann Institute of Science, Rehovot 76100, Israel}
    \author{Andrii V. Chumak}	
    \email{andrii.chumak@univie.ac.at}	
    \affiliation{Faculty of Physics, University of Vienna, 1090 Vienna, Austria} 
  	\author{Victor~S.~L'vov}
 	\email{victor.lvov@gmail.com}	
    \affiliation{Department of Chemical and Biological Physics, Weizmann Institute of Science, Rehovot 76100, Israel}
     
  \begin{abstract} 
  Recent advancements in spintronics and fundamental physical research have brought increased attention to the rare-earth-based magnetically ordered materials. One of the important properties of these materials is the temperature dependence of the spontaneous magnetization $M(T)$.  Recently, a successful framework was proposed for the theoretical description of $M(T)$ across the entire temperature range from $T=0$\,K to the Curie temperature $T_{_{\rm C}}$ in simple cubic ferromagnetics, EuO and EuS.   We extend this approach to
 compute and analyze  $M(T)$ for a more complex magnetic material: multi-sublattice collinear ferrimagnets such as   Yttrium Iron Garnet,  (YIG)  Y$_3$Fe$_5$O$_{12}$. YIG is a unique material that has a high Curie temperature well above room temperature.  It features a cubic crystallographic structure characterized by low anisotropy and minimal spin-orbit interaction. These properties result in an exceptionally low spin-wave damping, which is important for fundamental studies as well as engineering applications. The primary challenge in describing ferrimagnets theoretically arises from its intricate magnetic structure, characterized by multiple magnetic sublattices.
 We analyzed and generalized for multi-sublattice collinear ferrimagnets two well-known approximations describing $M(T)$. 
 The first approach is the Bloch $\frac32$-law, which describes the suppression of $M(T)$ due to spin-wave excitation, and is valid in the low-temperature limit $T\ll   T_{_{\rm C}}$. The second one is Weiss's mean-field approximation, which provides a reasonable description of  $M(T)$  near $T_{_{\rm C}}$.   Using a single tuning parameter, we combine these two approaches to describe $M(T)$ for any $ 0 \leq T \leq T_{_{\rm C}}$. The theoretical result for $M(T)$ aligns well with our measurements and the previously available experimental data across the entire temperature range. We also demonstrate that experimental and theoretical dependencies $M(T)$ follow the mean-field prediction $\sqrt{T_{_{\rm C}} -T }$ for almost all temperatures.  
\end{abstract}
\maketitle  

 \begin{figure*} 
\includegraphics[width=14 cm ]{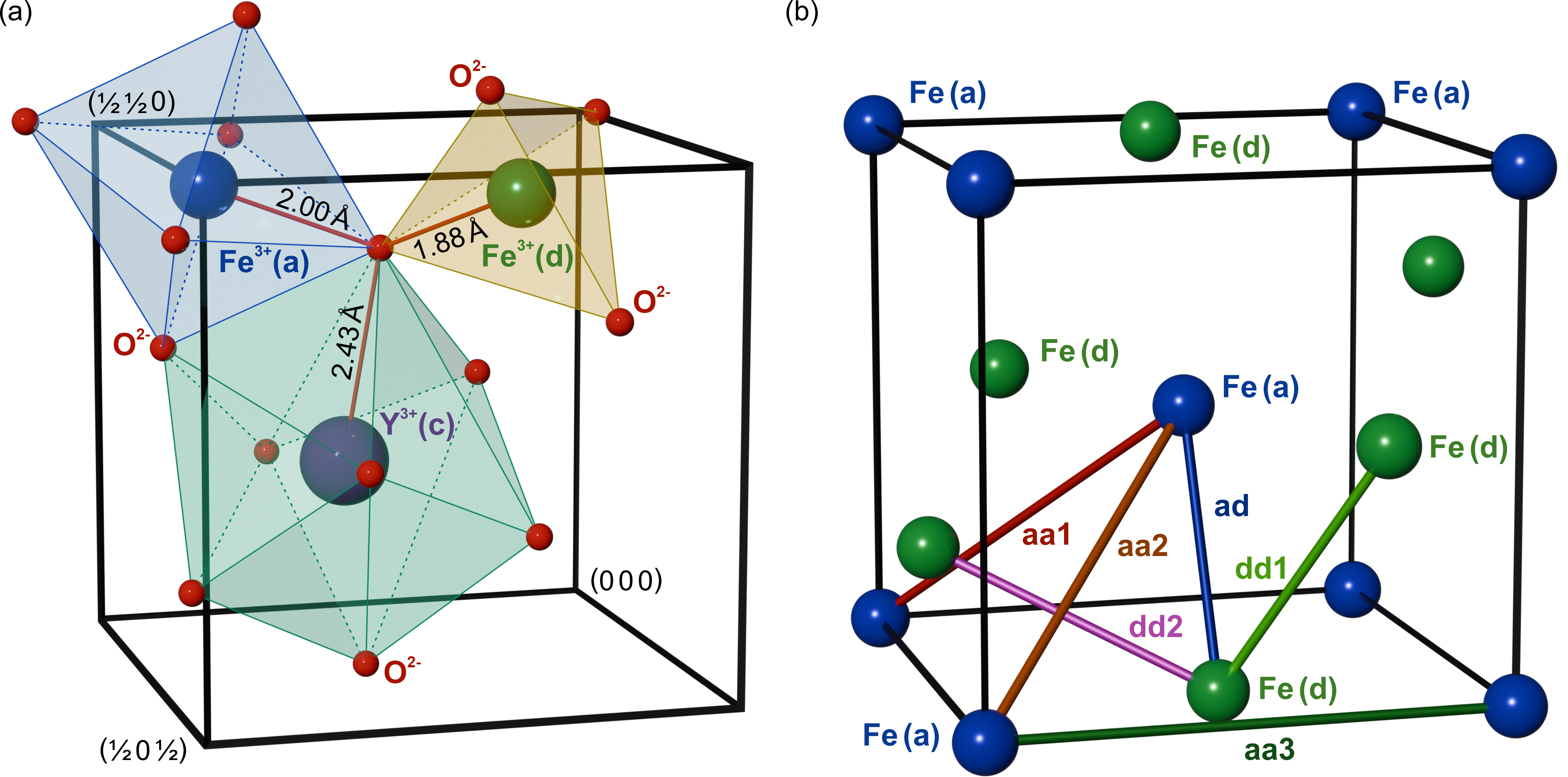}
  \caption{\label{f:1} 
 (a)  The crystallographic structure of  YIG  features a body-centered cubic (BCC) unit cell with the O$_{\rm h}$ symmetry group,  occupying half of a cube. This structure includes four formula units (Y$_3$Fe$_5$O$_{12}$)$\times 4$, amounting to a total of 80 atoms. Among these, there are 20 magnetic iron ions (Fe$^{3+}$) divided into two groups: 8 Fe$^{3+}$ ions occupying the (a) octahedral sites, represented as a blue ball and a blue-shaded area, and the remaining 12 Fe$^{3+}$ ions located in (d) tetrahedral sites, represented as a green ball and yellow-shaded area.
The dodecahedra (c) site occupied by Yttrium ions is shown as a black ball and green-shaded area, the  O$^{2-}$  ions are drawn in red.\\
    (b) The exchange pathways used in the effective Heisenberg Hamiltonian are labeled according to the sites they are connecting: $J_{\rm ad}$ as (ad), two different in symmetry (aa) interactions,  are denoted as $J_{\rm aa1}$ and $J_{\rm aa2}$, etc.} 
\end{figure*}

 \section{Introduction}  
Magnonics is a branch of spintronics that focuses on structures, devices, and circuits utilizing spin currents carried by magnons, which are the quanta of spin waves\, \cite{Chumak2015,Pirro2021,Flebus2024}. Similar to electric currents, magnon-based currents can be employed to carry, transport, and process information. The use of magnons enables the development of innovative wave-based computing technologies that are free from the limitations of modern electronics, such as energy dissipation caused by Ohmic losses\,\cite{Mahmoud2020,Wang2024}.  

Promising perspectives for engineering applications revived the attention to fundamental studies of the physics of magnetic materials.  To our understanding, the first step in developing a consistent theory of magnetic dielectrics is the study of their most fundamental properties: the temperature dependence of spontaneous magnetization $M(T)$ in thermodynamic equilibrium. 
The magnetization $M(T)$ plays a central role in the theories concerning the temperature dependence of other aspects of magnetodielectrics, such as magnetic susceptibilities,  heat capacities,  spin-wave frequency spectra, matrix elements of the magnetic dipole-dipole and exchange interactions, damping frequencies of magnons, and even quantum corrections to $M(T)$. 

Recently \cite{Kolokolov2025}, the behavior of $M(T)$ was studied for the simplest possible magnetic system: cubic ferromagnetic crystals containing only one magnetic ion per elementary cell. Currently, the only known and experimentally studied example of such a system is body-centered crystals of  EuO and EuS.

The analysis in \cite{Kolokolov2025} is based on the well-known Weiss mean-field approximation (MFA) \cite{Weiss1907}  with the Heisenberg exchange interaction \cite{Heisenberg1926}. In the MFA, an effective magnetic field acts on each magnetic ion due to exchange
interactions with its nearest neighbors. The larger their
coordination number $Z$ (i.e., the number of nearest neighbors), the smaller the fluctuations of this field, and the better the MFA performs.  In the limit $Z\to \infty$, MFA becomes exact. In EuO,
$Z = 12 \gg  1$ and the resulting value of the Curie temperature
for EuO  $T^{^{\rm MF}}_{_{\rm C}}\approx 
 86.6\,$K, predicted by MFA, is only about
20\% larger than its experimental value $T^{\rm Expt}_{_{\rm C}}\approx 
 69.8\,$K.

Significant improvements to the MFA were achieved 
by systematically accounting for the effects of spin waves on the fluctuations of the effective magnetic field and all $1/Z$-corrections to $M(T)$ using a specially developed perturbative diagrammatic technique for spin operators, without representing them through Bose operators. 
The resulting theory \cite{Kolokolov2025} is in excellent quantitative agreement with the experimental dependence of $M(T)$ for EuO and EuS
throughout the entire temperature range from $T = 0$ to the Curie temperature. In particular,  
the theoretical dependence $M(T)$ demonstrates a scaling behavior $M(T)\propto (T _{_{\rm C}} -T)^\beta$ with the scaling
index $\beta \approx 1/3$ in a wide range of temperatures, in agreement with the experimentally observed
apparent scaling in EuO and EuS.
 
A natural further step in the development of the theory of the spontaneous magnetization is to extend it to multi-sublattice ferri- and antiferromagnets. This is the objective of the present paper. To make our study more concrete and provide a basis for comparing our theory with experiments, we have chosen a specific example: the twenty-sublattice ferrimagnetic Yttrium Iron Garnet (YIG), Y$_3$Fe$_5$O$_{12}$.    YIG  is one of the most 
important materials in microwave magnetics.  In physics of magnetically ordered crystals, it plays a crucial role much like the role of Drosophila melanogaster (fruit fly) in genetics\,\cite{Rubin1988} or water in hydrodynamics \,\cite{Frisch1995}.
    
 YIG was first synthesized by Geller and Gilleo in 1957\,\cite{Geller1957}, resulting in a body-centered cubic structure shown in Fig.\,\ref{f:1}. 
Since then, it has helped to significantly advance our understanding of magnon dynamics. YIG has an exceptionally low magnon damping,
which allows magnon propagation over significant distances. This property makes it an ideal platform for developing microwave magnetic technologies. Such technologies have led to the creation of the magnon transistor and the first magnon logic gate, marking significant achievements in the field of magnonics\,\cite{Chumak2014,Serga2010,Levchencko2025,Dubs2020,Serha2025,Zenbaa2025}. The recent development by C. Dubs et al.~\cite{Dubs2020} of YIG film growth to a thickness as low as  10 nm using liquid phase epitaxy (LPE, \cite{Note1}) — a technique that provides the highest crystalline quality and, correspondingly, the lowest magnon damping  — is instrumental in this progress\,\cite{Chumak2022}.
 The primary reasons for the exceptionally low magnon damping in YIG are discussed in   Appendix\,\ref{A:dem}.

Despite extensive studies, the theoretical description of the magnetization in YIG, which accounts for the full complexity of its magnetic structure, was lacking.  The theory presented in this paper lays the groundwork for describing the magnetization temperature dependence in materials with nontrivial magnetic structures. 

 \begin{figure}  
 \includegraphics[width=0.99\columnwidth]{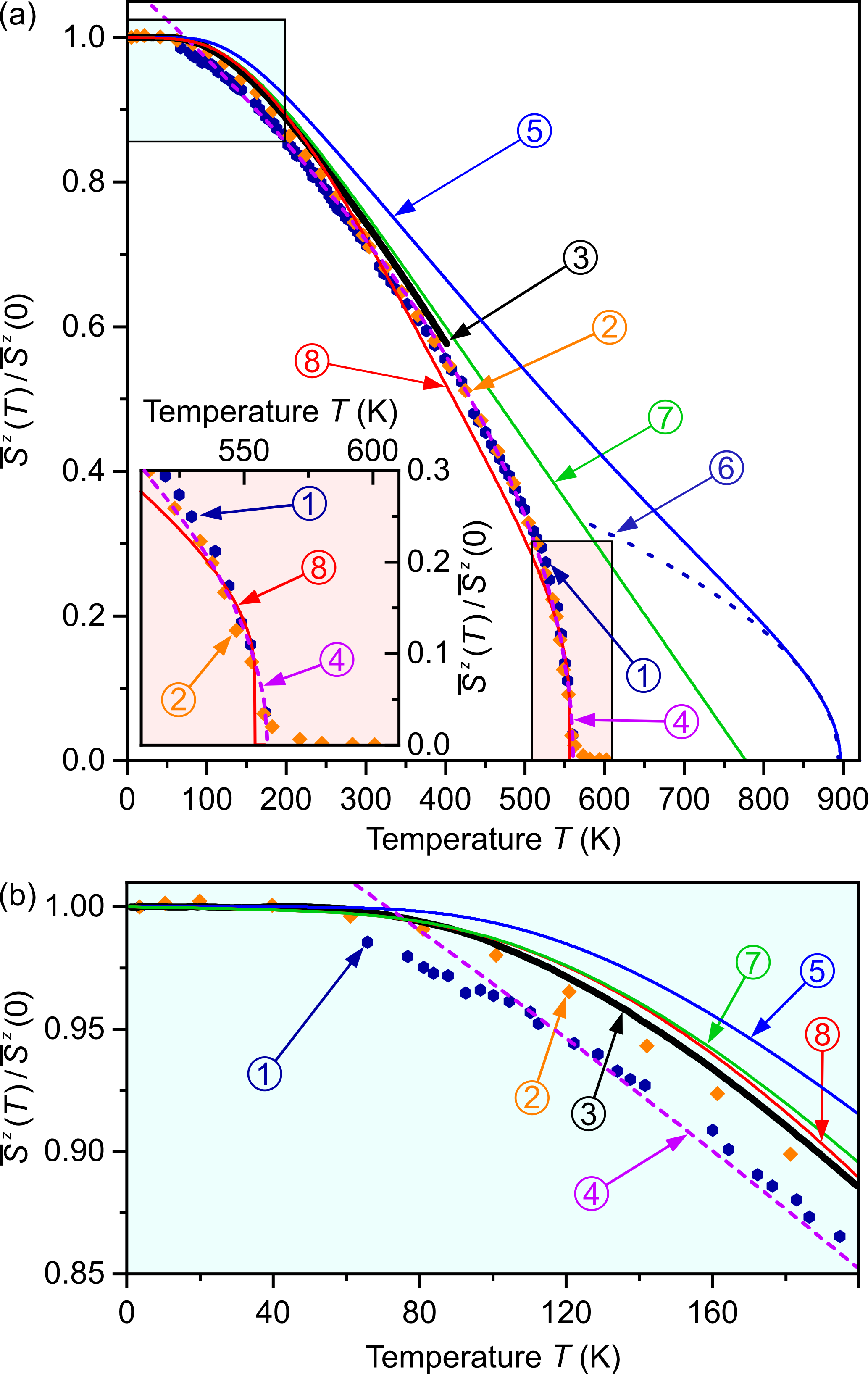} 
   \caption{\label{F:2}  (a) Temperature dependence of the normalized mean spin $\overline S^{z}(T)/ \overline S^{z}(0)$, equivalent to the normalized magnetization $M(T)/M(0)$, according to Eq.\,\eqref{Snorm}. Various experimental and theoretical results are labeled by circled numbers \textcircled{n}  of colors matching the color of the lines and described below. The close-up of the high-$T$ range marked by a rectangle with a pink background is shown in the inset. (b) The close-up of the low-$T$ range, marked by a rectangle with a light-blue background in the main panel (a). 
   \\   Various experimental datasets and theoretical results are labeled as follows.  Experimental data:   blue dots \textcircled{1} -- 1964-Anderson results \cite{Anderson1964},  orange dots \textcircled{2} -- 1974 data  Hansen and others \cite{Hansen1974}, black solid line \textcircled{3} -- our current results of a 1 mm YIG sphere magnetized in the ⟨111⟩ direction were measured via VSM.\\
Violet dashed line \textcircled{4} --  all-temperature fit of experimental data  \textcircled{2} $\overline{S}^{z}=A (T_{_{\rm C}}-T)^\beta$  with $A$~=~1.045 and $\beta \simeq  0.50\pm 0.005$; for details see Appendix \ref{A6}. \\
 Blue solid line \textcircled{5}  -- numerical solution of MFA Eqs.\,\eqref{4} and dashed blue line \textcircled{6} -- near-$T_{_{\rm C}}$ fit for MFA solution \textcircled{5}, $\propto \sqrt{T_{_{\rm C}}-T}$; \\ Green solid line \textcircled{7} -- numerical analysis of  Eq.\,\eqref{S-Td} for spin-wave (SW) approximation (SWA) to $\overline{S}^z(T)/ \overline{S}^{z}(0)$, exact at low temperature; Red solid line \textcircled{8}, -- the unified approach Eq.\,\eqref{26}, with the magnon frequencies $\omega_j(k,T)$ from \eqref{T-dep1} and $\delta=0.428$. 
 }
\end{figure}
\subsection*{\label{ss:plan}  Overview of the paper and main results}
Our aim in this paper is to improve the description of multi-sublattice ferrimagnets, focusing on   a theory of spontaneous magnetization $M(T)$ over
an entire temperature range from $T = 0$ to the Curie temperature $T_{_{\rm C}}$. 

The paper is organized as follows. Section \ref{s:History} provides a brief overview of the historical and physical background of the problem. We remind in Section\,\ref{ss:MFA} the Weiss MFA  based on the Heisenberg exchange Hamiltonian\,\eqref{Ham}, provide in Sec.\,\ref{ss:structure} the necessary information on the crystallographic and magnetic structure of  YIG, shown in Fig.\,\ref{f:1},  as well as the frequency spectra of spin waves in YIG (Sec.\,\ref{ss:spectra}).

 In Sec.\,\ref{s:MFA}, we analyze the existing theoretical approaches to the description of $M(T)$ and generalize them to the multi-sublattice magnetic structure. 
The theory is compared to existing experimental data \cite{Anderson1964,Hansen1974} and our measurements. Instead of using the normalized magnetization
$M(T)/M(0)$, which is assumed to be aligned with
the small external magnetic field $ \bm h (0, 0, h_{ z})$, we
adopt an approach more convenient  from the theoretical viewpoint to use the normalized  $\widehat{  z}$-projection of the mean
spin:
\begin{equation}\label{Snorm}
\frac{\overline S^{z}(T)} {\overline S^{z}(0)}  = \frac {M(T)}  {M(0)} \, .
\end{equation}
The mean spin $\overline S^{z}(T)$ is defined
as follows 
\begin{equation}\label{Sa}
  \overline S^{z}(T)= \frac 1 { N_{\rm lat}} \Big \langle \sum _j S^{z} _j \Big \rangle  \, .
\end{equation}  
Here   $j$ runs over magnetic ions lattice sites per unit volume and $N_{\rm lat} $ is their number.

We begin our study by presenting in  Fig.\,\ref{F:2} the earlier available experimental data\,\cite{Anderson1964,Hansen1974} for $\overline S^{z}(T)/\overline S^{z}(0)$, as blue and orange dots labeled as  \textcircled{1} and \textcircled{2} and our own accurate measurements down to temperatures near 2\,K, 
marked as the black solid line \textcircled{3}.  The experimental Curie temperature $T_{_{\rm C}}^{\rm Expt}\approx 560\,$K is clearly visible in the figure. A small tail above $T_{_{\rm C}}^{\rm Expt}$ originates from the external magnetic field (about 10\,kOe) used in experiments to suppress the domain structure of YIG, interfering with measurements.

The Landau's theory of the second-order phase transition\cite{landau1980}   and MFA\,\cite{Weiss1907}  as its particular realization, predict that the temperature dependence of $\overline S^{z}(T)$ for the temperatures close to  $T_{_{\text{C}}}$  exhibits the scaling behavior proportional to  $\sqrt{T_{_{\text{C}}} - T}$, which we will refer to as "normal scaling".  

We fit the experimental data with the $\overline S^{z}(T)\propto (T_{_{\text{C}}} - T)^\beta $ with $\beta=0.50 \pm 0.005$, shown in Fig.\,\ref{F:2} as a violet dashed line \textcircled{4}. A detailed description of the fitting method can be found in Appendix\,\ref{A6}. Unexpectedly, the normal scaling agrees well with experiments conducted not only near the critical temperature $  T_{_{\text{C}}} $ but across nearly all temperatures ranging from approximately $  T_{_{\rm C}}^{\rm Expt} \approx 560\,\text{K} $ down to about $  100\,\text{K} $, where $ \overline S^{z}(T)/\overline S^{z}(0) \approx 0.95 $. This behavior is in sharp contrast with the anomalous scaling of $  M(T) $ observed in simple ferromagnets like EuO and EuS, where the exponent $  \beta$  is approximately $  1/3 $ \cite{Kolokolov2025}. 

These two facts --- (i) the normal scaling covers most of the temperature range $M(T)$ in YIG, and (ii) the apparent scaling $M(T)$ in cubic one-sublattice ferromagnets with $Z=12$ is anomalous --- clearly indicate that there is more in the multi-sublattice magnetic crystals than meets the eye. A generalization of  MFA with an explicit account of the complex magnetic structure is in order. 

To this end, in Sec.\,\ref{ss:background},  we consider the  MFA  for two-sublattice ferrimagnets and derive Eq.\,\eqref{4} that describe it.  
Next, we derived an analytical expression for the Curie temperature, Eq.\,\eqref{5A}, along with the ratio of the sublattice magnetizations near $T_{_{\rm C}}$, Eq.\,\eqref{5B}.
  The numerical solution of Eq.\,\eqref{4}  in the form  Eq.\,\eqref{Sz}   is plotted in Fig.\,\ref{F:2} by the blue solid line \textcircled{5}.
As expected, near $T_{_{\rm C}}^{^{\rm MF}}$ it demonstrates normal scaling behavior $  \overline S^{z} (T)\propto \sqrt{T_{_{\rm C}}^{^{\rm MF}}-T}$, denoted  by the blue dashed line \textcircled{6}.

 The resulting Curie temperature $T_{_{\rm C}}^{^{\rm MF}}\approx 894\,$K calculated using values of the YIG exchange integrals \cite{Cherepanov1993} exceeds its experimental value $T_{_{\rm C}}^{\rm Expt}=560\,$K almost twice, while in EuO the discrepancy is only about 20\%. 
 This is discouraging but not unexpected, since the dominant inter-sublattice interactions in YIG are carried through only $Z_{\rm YIG}\approx 5 $  neighbors on average. Therefore, the applicability parameter $1/Z_{\rm YIG} \approx 1/5$ is larger than $  1/Z_{\rm EuO} \lesssim 1/12$, and poorer performance of MFA compared to EuO is expected.   
 
The second problem with the MFA, common for EuO and YIG, is that it does not account for the transverse fluctuations of the neighboring spins $\bm S_j$, which influence the given spin $\bm S_i$. 

 This issue is addressed in Sec. \ref{ss:low-T}  where we formulated spin-wave approximation (SWA)  \eqref{S-Td} for suppression of magnetization in multi-sublattice magnetics. To calculate the magnitization in the framework of SWA, an explicit knowledge of the magnon frequency spectra is required.
 
 It is well known that the frequency spectra of the magnetic material with  $n$ sublattices involve $n$ branches. 
 For collinear magnetic structures with $n_+$ sublattices oriented along the external magnetic field and $n_-=n-n_+\leq n_+$ sublattices oriented in the opposite direction,  the total $n$ branches can be categorized into two groups: $n_+$ ``ferromagnetic" modes and $n_-$ ``antiferromagnetic" modes.  According to Eq.\,$ \eqref{S-Td}$, the excitation of a single ferromagnetic magnon reduces the total $z$-projection of spin, $\overline S^z$,  by unity.  
In contrast, the excitation of a single antiferromagnetic magnon increases $\overline S^z$ by unity.

In YIG, there are $n_+=12$ ferromagnetic modes and $n_-=8$ antiferromagnetic modes, making a total of $n=20$ frequency modes, as is shown in Fig.\,\ref{F:3}. The ferromagnetic modes are represented by green lines,  while the antiferromagnetic modes are shown by blue lines.  

 Formally speaking, Eq.\,\eqref{S-Td} is exact in the low-temperature range, say for $T\ll T_{_{\rm C}}$.  To our surprise, the numerical solution of Eq.\,\eqref{S-Td}, represented by the green solid line \textcircled{7} in Fig.\,\ref{F:2}, aligns well with the experimental data \textcircled{1}, \textcircled{2}, and \textcircled{3} from $ T = 0 $ up to $ T \approx 400\,$K. At higher temperatures  $\overline S^{z} (T)$ decreases almost linearly, crossing zero  at 
$T_\times\approx 770\,$K and then formally becomes negative. 

 In Eq.\,$ \eqref{S-Td}$, we assumed temperature-independent frequency dispersion relations taken at $T=0\,$K. However, the frequency spectra depend on $T$  according to Eq.\,\eqref{T-dep}, where a number of additional assumptions were made. When this temperature dependence
 is taken into account in Eq.\,\eqref{S-Td}, the resulting numerical solution, which is still correct at low $T$, is very strongly suppressed with the increasing temperature, crossing now zero at $T_\times \approx 410\,$K.   
 
 At this point, we have an MFA solution, having a correct behavior near $T_{\rm _C}$ but overestimating $M(T)$, and a SWA solution that has a correct low-$T$ behavior but underestimates the magnetization for larger temperatures. A natural way forward is to combine these two approaches.

 In the simple case of a one-sublattice ferromagnet, this concept was demonstrated in Ref.\,\cite{Kolokolov2025} 
 and briefly reminded in Sec.\,\ref{ss:EuO} for completeness. Its generalization for multi-sublattice ferrimagnets like YIG is not straightforward and requires significant efforts, including the application of the diagrammatic perturbation approach for spin operators, such as the Belinicher-Lvov diagrammatic technique\,\cite{Belinicher1984}. This approach is beyond the scope of the current paper.

Instead, in Sec.\,\ref{ss:YIG}, we propose a simplified version of this unification, referred to below as the "SW-MFA minimal model".
This model incorporates Eq.\eqref{26} for $\overline S^{z}$ and includes a tuning parameter $\delta$ that regulates the temperature dependence of the magnon frequencies as described by Eq.\eqref{T-dep1} in Sec.\,\ref{sss:T-dep}.
To choose the parameter $\delta$, we require that the resulting  $T_{\rm _C}$, given by Eq.\,\eqref{Tc}, coincides with its experimental value.

  The numerical solution of  Eqs.\,\eqref{T-dep1}  and \eqref{26} with $\delta=0.428$ is shown in Fig.\,\ref{F:2} as the red solid line \textcircled{8}. Bearing in mind that we computed the solution \textcircled{8} using the single tuning parameter $\delta$, we consider its agreement with experimental data for $\overline S^{z}(T)$ across the entire temperature range as fully satisfactory.
   It is worth noting that the theoretical dependence \textcircled{8} of $\overline S^{z}$ retains normal scaling over $100 \lesssim T \lesssim T_{\rm _C}$, much wider than the MFA.

As previously discussed, the method for studying the temperature dependence of magnetization in single-sublattice ferromagnetics like EuO and EuS, as presented in \cite{Kolokolov2025}, shares several similarities with the approach used for multi-sublattice ferrimagnets, such as YIG, developed in this paper. While the initial steps are similar, they diverge significantly in the final stages. Details of their comparison are provided in Appendix \ref{s:comp}.

We summarize our findings in Sec.\ref{s:sum}.


\section{\label{s:History} Historical and physical background} 
In this section, we will provide the reader with general physical and historical information 
 necessary for completeness of the presentation. 

 \subsection{\label{ss:MFA} Weiss-Heisenberg mean-field approximation}
 The analysis of magnetic properties in magnetodielectrics begins with the Weiss mean-field approximation\,\cite{Weiss1907} (MFA) with the Heisenberg exchange Hamiltonian\,\cite{Heisenberg1926}. In our paper, we adopt the simplest form of this Hamiltonian, which describes the interaction between the spins $ \bm S_i$and $ \bm S_j$, located at positions $ \bm r_i$ and $ \bm r_j$ respectively: 
  \begin{equation}\label{Ham}
{\cal H}= -\frac12 \sum _{i\ne j}J_{ij} \bm S_i \cdot \bm S_j= -\sum _{i>j}J_{ij} \bm S_{i} \cdot \bm S_j \ .
\end{equation} 
Here, exchange integrals $ J_{ij}$ represent the energy associated with the exchange interactions. In addition to exchange interactions, much weaker magnetic dipole-dipole interactions may also play some role in the very vicinity of the Curie temperature \,\cite{Fisher1973,Aharony1973,Kornblit1975}.
The dipole-dipole interactions determine the frequency spectrum of long spin-waves. We discuss their contribution in Appendices \ref{D} and \ref{E}.

Weiss  MFA  ignores fluctuations of all spins, replacing their actual, time-dependent values $\bm S_j(t)$ by the mean value $\bm S_j(t)\rightarrow \overline S^{z}$.  If so, the effective magnetic field (in temperature units), acting on every spin,  is given by  
 \begin{subequations}
\begin{equation}\label{heff}
H^{\rm eff}= \overline S^{z} {\cal J}_0\,, \quad {\cal J}_0=\sum _j J_{ij}\ .
\end{equation} 
According to the quantum statistics \,\cite{landau1980}, the mean value $\overline S^{z}$ of some spin in a magnetic field $H$ at a temperature $T$
is given by the normalized Brillouin function for the magnetic ion spin $S$ (for example, in EuO $S=7/2$, while for YIG $S=5/2$)
\begin{eqnarray}  \label{bs}
 ~ \hskip - .5 cm b_{_S}(x)&=& \Big ( S+\frac12 \Big )\coth \! \Big [ \Big  (   S+\frac12    \Big )x\Big ]- \frac 1{2}
\coth\Big  ( \frac{x}{2 }\Big ) \    .\label{Bri}
 \end{eqnarray} 

 Therefore, the basic equation for $\overline S^{z}$ in the MFA takes the simple form \cite{Kolokolov2025}.
 \begin{equation}\label{MFA}
 \overline S^{z}(T)= b _{_S}\Big[\frac{\overline S^{z}(T) {\cal J}_0}{T}\Big] \ .
 \end{equation}
 \end{subequations}
Here ${\cal J}_0$ is defined by Eq.\,\eqref{heff}. For nearest-neighbor  interactions with coordination number $Z$, ${\cal J}_0$  is given by ${\cal J}_0 = Z J$, where $J$ represents the exchange integral.  
 
 Expanding $b_{_{\rm S}}(x)$ for small $x$
 \begin{equation} \label{expB}
b_{_S}(x)=  \frac{S(1+S)x}{3   }  + \frac {x^3 } {720   }[1- (1+2 S)^4]+ \dots\, 
\end{equation}
 one finds the  Curie temperature in the MFA:
 \begin{equation}\label{3} T _{_{\rm C}}^{^{\rm MF}}= \frac{S(S+1)  {\cal J}_0}{3} \ , 
 \end{equation}
at which the average spin $ \overline{S}^{z}$vanishes.

\subsection{\label{ss:structure} Crystallographic and magnetic structure of YIG}

The first obstacle in our pursuit is YIG's complicated crystallographic structure.
 The unit cell of YIG includes four formula units (Y$_3$Fe$_5$O$_{12}$)$\times 4$, i.e., 80 atoms.   The lattice has a body-centered cubic (BCC) unit cell of O$_{\rm h}$ symmetry group with a lattice constant of $a=12.4\times 10^{-8}$ cm, occupying half of a cube as shown in Fig.\,\ref{f:1}. 
 Magnetic Fe$^{3+}$ ions  have spin $S=\dfrac 52 $ and magnetic moment $\mu= 5 \, \mu_{_{\rm B}}$, where $\mu_{_{\rm B}}$ is the Bohr magneton.  They occupy two inequivalent positions with regards to the symmetry of their immediate O$^{2-}$ environment -  $N_{\rm a}=8$  octahedral sites (a), shown in Fig.\,\ref{f:1} as a blue ball and a light-blue-shaded area, and  $N_{\rm d}=12$  tetrahedral sites (d), shown as a green ball and a yellow-shaded area.   Explicit coordinates of magnetic Fe$^{3+}$ ions are listed in Appendix \,\ref{ss:A1}.

\begin{figure} 
\includegraphics[width=1\columnwidth]{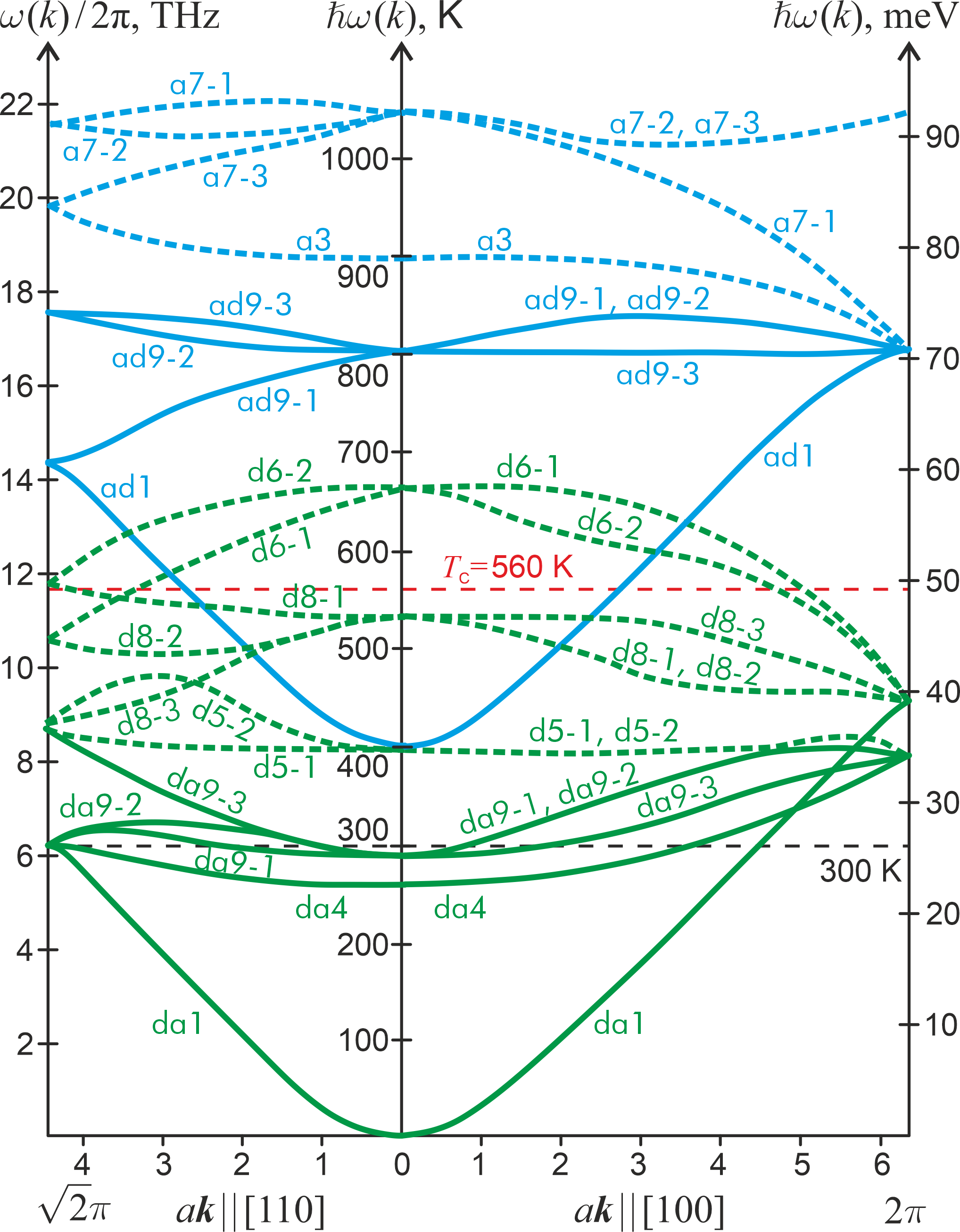}
\caption{\label{F:3} The plots of an analytical result of Ref. \cite{Cherepanov1993} for all 20 branches of the magnon energy spectra in YIG for $\bm k \| [110]$ (left part of the $\bm k$ axis and for $\bm k\| [100] $    (right part of the $\bm k$ axis).  Three vertical axes represent various ways to quantify the magnon energy and frequency. The modes corresponding to (a) -interactions are shown as dashed blue lines; the  (d)-modes are plotted as dashed green lines; the (ad)-modes are shown as solid blue lines, and the (da)-modes are shown as solid green lines. Horizontal black dashed line indicates the room temperature $T=300$ K, and the horizontal red dashed line marks the  YIG Curie temperature $T_{_{\rm C}}\approx 560\,$K.  These spectra are quite similar to recent results by Princep et al. \cite{Princep2017a,Princep2017j} as listed in  Tab.\,\ref{t:2}. 
}
\end{figure} 


 Furthermore,
  the sites of type (a) have  $Z_{\rm aa} = 8$  nearest neighbors of type (a) and  $Z_{\rm ad} = 6$  nearest neighbors of type (d). On the other hand, the sites of type (d) have  $Z_{\rm dd} = 4$  nearest neighbors of type (d) and  $Z_{\rm da} = 4$  nearest neighbors of type (a). Therefore,  $Z_{\rm ad} \neq Z_{\rm da}$.
   The sites also vary in the configurations of neighboring O$^{2-}$ ions, which create a superexchange interaction between the magnetic ions.  
 
Over several decades, a variety of exchange integrals for YIG have been proposed, some of which are listed in Tab.\,\ref{t:1} and discussed in Appendix\,\ref{ss:J}. For this paper, we adopt the following values:
\begin{equation}\label{int}
J_{\rm ad}= -79.6\,\mbox{K}\,, \ J_{\rm aa}= -7.6\,\mbox{K}\,, \ J_{\rm dd}= -26.8\, \mbox{K}\,,
 \end{equation}
 suggested in Ref.\,\cite{Cherepanov1993}. We see that the leading interaction is the antiferromagnetic (a-d) interaction characterized by negative exchange integrals  $J_{\rm ad} = J_{\rm da} < 0 $. This results in a collinear antiferromagnetic arrangement, where all spins in (d)  positions $\bm{S}_{\rm d}$ are parallel to each other and oriented  
 along the  positive   $\widehat{\bm{z}}$-direction, while all (a)-spins  $\bm{S}_{\rm a}$  are also parallel but oriented 
 along the  negative $\widehat{\bm{z}}$-direction.

 The overall average spin within the elementary cell of YIG is a  sum of the  mean spins for each of the magnetic sites, as defined by Eq.\,\eqref{Sa} 
  \begin{equation}\label{Sz}
   \overline S^{z}(T) =  N_{\rm d}  \overline S^{z}_{\rm d}(T) + N_{\rm a} \overline S^{z}_{\rm a} (T) \ .
 \end{equation}  

 Negative values of relatively small exchange integrals $J_{\rm aa}$ and $J_{\rm dd}$ create frustration within sublattices (a) and (d), which is overcome by the dominant  $J_{\rm ad}$ coupling. Therefore, they cannot disrupt the intrinsic ferromagnetic arrangements inside of (a) and (d) sublattices. 

\subsection{\label{ss:spectra} Spin-waves  spectra of YIG}
To analyze the effect of spin waves (magnons) on the temperature dependence of magnetization in  YIG, we need to determine their frequency spectra   $\omega_j(\bm{k})$. A standard approach for finding   $\omega_j(\bm{k})$ involves considering spin operators in terms of boson operators for creation and annihilation,  $a$ and $a^\dag$. To achieve this, Harris \cite{Harris1967} employed the well-known Holstein-Primakoff representation \cite{Holstein1940} within a linear approximation.
\begin{subequations}\label{HP}
\begin{align} \label{HPa}
S^{z}_j=& \mp [S- a^\dag_j a_j]\,, \\ \label{HPb}
S^\pm_j = \,\,& S^x \pm i S^y= \sqrt {2S}\, a^\dag_j\,, \\ \label{HPc}
 S^\mp_j = \,\,& S^ x_j \mp i S^y= \sqrt {2S} \, a _j\,,
 \end{align} \end{subequations}
 where we take the upper choice of sign for the (a) sites and the lower choice of sign for the (d) sites. This allows us to specify $\bm S_j$ in Eqs.\,\eqref{Ham}. 

 After   the Fourier transform from  $a_j\equiv a(\bm r_j )$ to  $a_j(\bm k )$, the Hamiltonian \eqref{Ham} takes the quadratic form diagonal in $\bm k$,  $\displaystyle  {\cal H}_2=\sum _{\bm k}H(\bm k)$, where 
 \begin{align} \begin{split} \label{Ham2A}  
H(\bm k)=& \sum_{i,j=n_++1}^ {n } A_{ij}(\bm k) a_i^*(\bm k) a_j(\bm k)\\ &+
\sum_{i,j=1}^ {n_+} D_{ij}(\bm k) a_i^*(\bm k) a_j(\bm k) \\
& + \sum_{i=n_++1}^ {n} \sum_{j=1}^ {n_+}\big [ B_{ij}(\bm k) a_i^*(\bm k) a_j^*(-\bm k)
+ \mbox{c.c} \big ] \,,   
\end{split}\end{align}
Recall that in YIG $n_+=12$, $n_-=8$, and $n=n_++n_-=20$.  The explicit and quite cumbersome equations for $A_{ij}(\bm k)$, $B_{ij}(\bm k)$ and $D_{ij}(\bm k)$ one finds in \cite{Cherepanov1993}.

A quadratic Hamiltonian \eqref{Ham2A} can be diagonalized using the Bogolyubov linear $(u,v)$-transformation. In our case, the transformation takes a simpler form without certain terms  \cite{Cherepanov1993}:
 \begin{align}  \nonumber 
a_i(\bm k)=& \sum_{j=n_++1}^{n} u_{ji}^{(1)}(\bm k) b_j (\bm k) - \sum_{j=1}^{n_+} v_{ji} (\bm k) b_j^* (-\bm k)\,, \\ \label{uv}  i=& n_++1, \ \dots\ ,n\,,\\  \nonumber
a_i^*(-\bm k)=& \sum_{j=n_++1}^{n} v_{ji}(\bm k) b_j (\bm k) + \sum_{j=1}^{n_+} u_{ji}^{(2)} (\bm k) b_j^* (-\bm k)\,, \\  \nonumber i=& 1, \ \dots\ , n_+ \ .
 \end{align}
In the new variables, the Hamiltonian \eqref{Ham2A} takes the diagonal form
\begin{equation}\label{10}
H_{\bm k}= \sum_{j=1}^{n} \omega_j(\bm k) b_j(\bm k)b_j^*(\bm k)\ .
 \end{equation} 

  YIG contains $n=20$ magnetic ions in its elementary cells, which results in 20 branches of the magnon spectra, according to Eq.\,\eqref{10}. As shown in Ref.\,\cite{Cherepanov1993}, they can be divided into three groups based on the nature of the oscillations for $k=0$.

There are four branches in which only the spins at the (a) positions oscillate at $k = 0$. These are indicated in Fig.\,\ref{F:3} by blue dashed lines. In eight additional branches, shown by green dashed lines, only the spins at the (d) positions oscillate at $k = 0$. In the remaining eight branches, the spins oscillate in both the (a) and (d) positions.

These branches can be divided into two groups, each containing four modes, referred to as the (da)- and (ad)-modes, as illustrated in Fig.\,\ref{F:3} with solid green and blue lines. The 
notation (da) is assigned to the four ferromagnetic (FM) modes. We will show below that excitation of a magnon of this type decreases  $z$-projection of the total spin $S^z$ by unity, similar to the effect produced by the excitation of any (d)-branch magnon. In contrast, the excitation of a magnon from the four antiferromagnetic (AFM) branches, labeled (ad) (which includes the AFM mode ad1), increases $S^z$ by unity. Any (a) mode shares this property.

As shown in Fig.\,\ref{F:3}, at temperatures below approximately 200 K, only the  FM branch  (da1) originating from a one-dimensional identical representation becomes excited. This significantly simplifies the analysis of sublattice magnetization, especially at relatively low temperatures.

\section{\label{s:MFA} Mean-field approximation for YIG }
In this section, we analyze the mean-field approximation for YIG, consider the magnon contribution to the temperature dependence of $\overline{S}^{z}$, and compare these results with those available in the literature and our measurements.

 \subsection{\label{ss:background}MFA for two-sublattice ferrimagnets} 
As discussed in the Introduction, the MFA focuses solely on the mean values of the spin $\overline S^{z}_j$ at site $\bm r_j$. 
In YIG, there are only two crystallographically distinct positions of Fe ions, labeled (a) and (d), with mean spins $ \overline S^{z}_{\rm a}(T)$  and $\overline S^{z}_{\rm d}(T)$. As a result, within the framework of MFA, YIG can be regarded as a two-sublattice ferrimagnet.

 In a manner similar to Eq.\,\eqref{heff}, the effective magnetic fields acting on spins (a) and (d) can be expressed as follows:
  \begin{align}\begin{split}\label{HeffA}
H^{\rm eff}_{\rm a}&=\overline S_{\rm a}^{z} {\cal J}_{\rm aa}+\overline S_{\rm d}^{z}  {\cal J}_{\rm ad}\,, \ 
H^{\rm eff}_{\rm d} =\overline S_{\rm a}^{z} {\cal J}_{\rm da}+\overline S_{\rm d}^{z}  {\cal J}_{\rm dd}\ .
 \end{split}\end{align}
Here ${\cal J}_{\rm aa}\equiv J_{\rm aa}Z_{\rm aa}$, ${\cal J}_{\rm dd}\equiv J_{\rm dd}Z_{\rm dd}$, and  ${\cal J}_{\rm ad}\equiv J_{\rm ad}Z_{\rm ad}$,   ${\cal J}_{\rm da}\equiv J_{\rm ad}Z_{\rm da}$.  Here, we recall that $Z_{ij}$ are   the coordination numbers $Z_{\rm ad}=6$, $Z_{\rm da}=4$, $Z_{\rm aa}=8$, and $Z_{\rm dd}=4$. When combined with the   experimental values of the exchange integrals \eqref{int}, the resulting values are
\begin{align}\begin{split}\label{cal-J}
{\cal J}_{\rm ad}\approx& -478\,\mbox{K}, \quad {\cal J}_{\rm da}\approx -318\,\,\mbox{K} , \\ 
{\cal J}_{\rm aa}\approx& -\ 61\,\,\mbox{K} , \quad  {\cal J}_{\rm dd}\approx -107\,\,\mbox{K} \ . 
\end{split}\end{align}

Equations \eqref{HeffA} lead to  a natural generalization of the mean  field equation \eqref{MFA}
\begin{subequations}\label{4}
\begin{align} \label{4A}
\overline S_{\rm a}^{z}(T)=\,\,&   b _{_S}\Big[ \frac{\overline S_{\rm a}^{z} (T){\cal J}_{\rm aa}+\overline S_{\rm d}^{z} (T) {\cal J}_{\rm ad} }{T}\Big]\,,  \\ \label{4B} 
\overline S_{\rm d}^{z}(T)=\,\,& b _{_S}\Big[ \frac{\overline S_{\rm a}^{z}(T) {\cal J}_{\rm da}+\overline S_{\rm d}^{z}(T)  {\cal J}_{\rm dd} }{T}\Big] \ .
 \end{align}\end{subequations}
 
Employing the expansion   \eqref{expB},  we solve $\overline S_{\rm a}^{z}=\overline S_{\rm d}^{z}=0$, to find the Curie temperature $T _{_{\rm C}}^{^{\rm MF}}$ within the MFA
\begin{subequations}\label{5}
\begin{align}\begin{split}\label{5A}
 T _{_{\rm C}}^{^{\rm MF}}=\,& \frac{S(S+1)}{6}\big [{\cal J}_{\rm dd} +{\cal J}_{\rm aa} \\ 
& + \sqrt{ ({\cal J}_{\rm aa} -{\cal J}_{\rm dd})^2 + 4{\cal J}_{\rm ad}{\cal J}_{\rm da}} \big ] \ .
\end{split}\end{align}
 We also found a ratio  $  \overline S_{\rm d}^{z} /\overline S_{\rm a}^{z} $  when $T$ approaches $T _{_{\rm C}}^{^{\rm MF}}$ from below:
\begin{align}\begin{split}\label{5B}
\frac{\overline S_{\rm d}^{z} } {\overline S_{\rm a}^{z}}=\,& \frac{1}{2 {\cal J}_{\rm ad} }\big [{\cal J}_{\rm dd} -{\cal J}_{\rm aa}\\
& + \sqrt{ ({\cal J}_{\rm aa} -{\cal J}_{\rm dd})^2 + 4{\cal J}_{\rm ad} {\cal J}_{\rm da} } \big ]. 
\end{split}\end{align} \end{subequations}
In the limits ${\cal J}_{\rm ad}\gg {\cal J}_{\rm aa}, {\cal J}_{\rm dd}$ Eq.\,\eqref{5A} takes a simple form,
\begin{subequations}\label{6}
\begin{eqnarray}\label{6A}
  T_{_{\rm C}}^{^{\rm MF}} \approx  \frac{S(S+1)}{3}\sqrt { {\cal J}_{\rm ad}  {\cal J}_{\rm da} }  \,,
\end{eqnarray}
similar to Eq.\,\eqref{3}. Then, the ratio \eqref{5B} becomes     
\begin{align} \label{6B}
\frac{\overline S_{\rm d}^{z} } {\overline S_{\rm a}^{z}}=   \sqrt {\frac { {\cal J}_{\rm da}}{{\cal J}_{\rm ad} } }=\sqrt{\frac 23}\approx 0.82 \ .
 \end{align}\end{subequations}

 One can show  that Eqs.\,\eqref{4} results in the square-root scaling of $\overline S_{\rm d}^{z}$ and $\overline S_{\rm a}^{z}$ as $T$ approaches $T_{_{\rm C}}^{^{\rm MF}}$ from below
\begin{equation}\label{below}
 \overline S_{\rm a}^{z}\propto \sqrt{T _{_{\rm C}}^{^{\rm MF}} -T}\,, \quad \overline S_{\rm d}^{z}\propto \sqrt{T _{_{\rm C}}^{^{\rm MF}} -T}\,,
 \end{equation} 
 similar to any model that uses  MFA\,\cite{landau1980}.
 
 Consider now the solution of Eqs.\,\eqref{4} when the temperature approaches zero from above $T\to 0^+$. In this limit, the argument of $b_{_S}$ in \eqref{4A} approaches $-\infty$ (assuming that $\overline S_{\rm d}^{z} >0$, $J_{\rm ad}<0$, and $|J_{\rm ad}| \gg |J_{\rm aa}|$) and approaches $+\infty$ in \eqref{4B} (if $\overline S_{\rm a}^{z} <0$ and $|J_{\rm ad}| \gg |J_{\rm dd}|$). Using the equation for $b_{_S}(x)$
in  the limits $|x|\gg 1$,
\begin{subequations}\label{exp}
\begin{align}\begin{split}\label{expA}
b_{_S}(x) &\approx   S - \exp (-x)\,, \quad x\to + \infty\,,\\
b_{_S}(x) &\approx   -S +  \exp (x)\,, \quad x\to - \infty\,,
\end{split}\end{align}  
 we find that 
\begin{align}\begin{split} \label{smallT} 
 \overline{S}_{\rm a}^{z} \approx & -S + \exp \Big [  \frac{S ({\cal J}_{\rm ad}- {\cal J} _{\rm aa})}{T}\Big]\,,   \\ 
  \overline{S}_{\rm d}^{z} \approx & + S -     \exp \Big [  \frac{S ({\cal J}_{\rm  da}- {\cal J} _{\rm dd})}{T}\Big]   \ .
 \end{split}\end{align} \end{subequations}
We see that as  $T\to 0$, 
$\overline{S}_{\rm a}^{z}$   approaches  $- S $,  while  $\overline{S}_{\rm d}^{z}$   approaches  $+ S $.  At small $T$, the deviation of  both $|\overline{S}_{\rm a}^{z}|$  and $|\overline{S}_{\rm a}^{z}|$  from $S$ is exponentially small.
 
\begin{figure}[t] 
      \includegraphics[width=0.95\columnwidth]{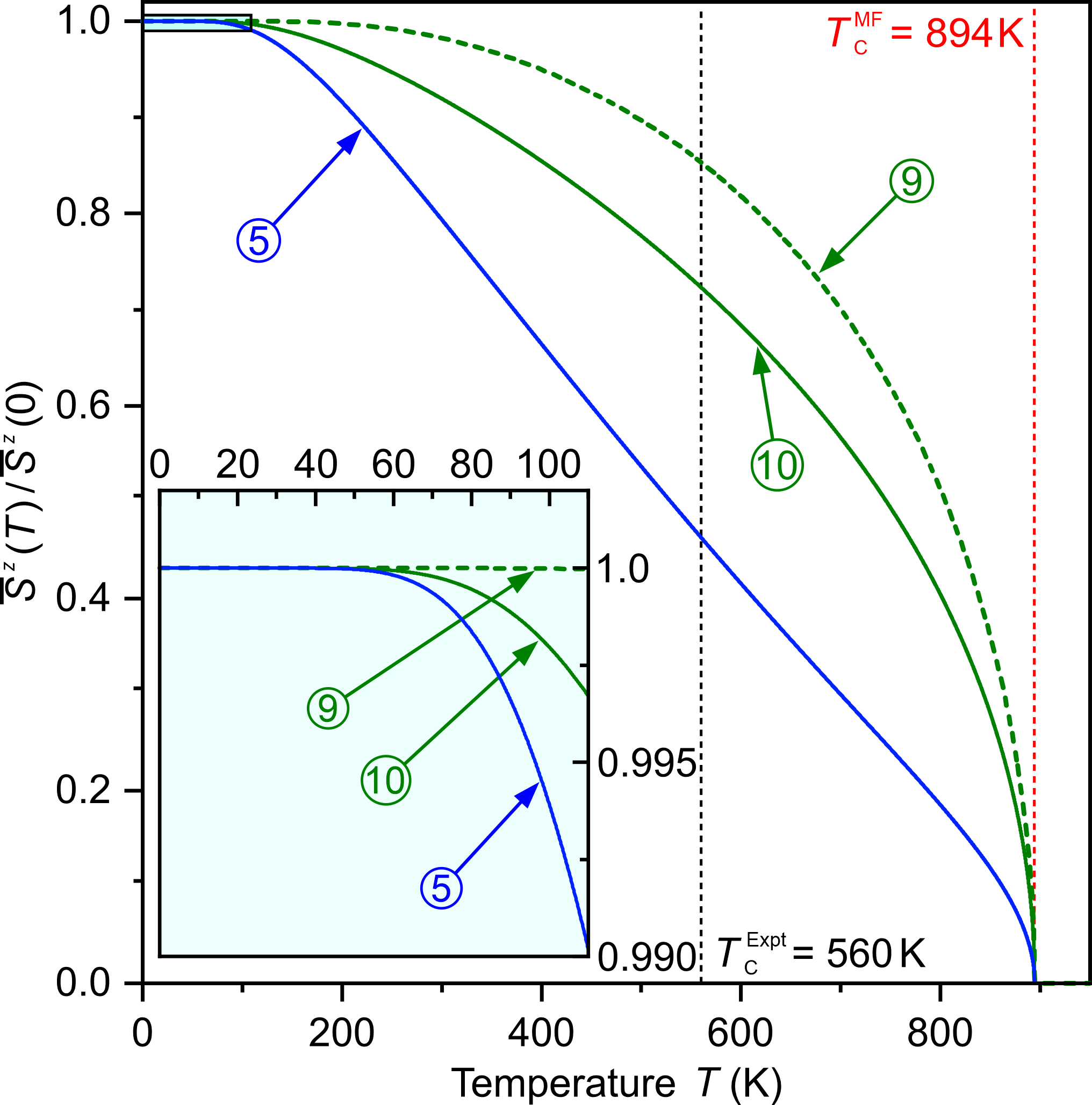}  
\caption{\label{F:4} The numerical solution for the normalized mean spin $\overline S^{z}(T)/ \overline S^{z}(0)$. The results of numerical solutions are denoted as follows:\\ 
 Solid blue line \textcircled{5} (the same as in Fig.\,\ref{F:2})-- the normalized total spin, defined  by Eq.\,\eqref{Sz}, \\
Green dashed and solid lines \textcircled{9} and \textcircled{10} represent the normalized spins of    sublattice (a) $-\overline S^{z}_\mathrm{a}(T)/\overline S^{z}_\mathrm{a}( 0)$, and sublattice (d) $\overline S^{z}_\mathrm{d}(T)/\overline S^{z}_\mathrm{d}( 0)$ respectively, defined by Eqs.\,\eqref{4} based on exchange integrals from \cite{Cherepanov1993}. 
Vertical dashed lines mark the Curier temperatures $T_{_{\rm C}}^{\rm Exp}$ (black) and $T_{_{\rm C}}^{\rm MF}$ (red).\\
Inset: high-resolution close-up of the low-$T$ range marked in the main panel as a rectangle with light-blue background. 
 }
\end{figure}
Fig.\,\ref{F:4} presents the results of the numerical solutions of MF Eqs.\,\eqref{4}. 
Solid blue line \textcircled{5}, also shown in Fig.\,\ref{F:2}, represents the
normalized total spin in YIG, defined by Eq.\,\eqref{Sz} with $N_{\rm d}=12$ and  $N_{\rm a}=8$. The green dashed and solid lines \textcircled{9} and \textcircled{10} correspond to the normalized spins of the sublattice (a) $-\overline S^{z}_\mathrm{a}(T)/\overline S^{z}_\mathrm{a}( 0)$, and sublattice (d) $\overline S^{z}_\mathrm{d}(T)/\overline S^{z}_\mathrm{d}( 0)$, respectively. Here, in \eqref{4} we used the exchange integrals \eqref{cal-J}. 

  The resulting numerical Curie temperature $T_{_{\rm C}}^{^{\rm MF}}=894\,$K agrees with Eq.\,\eqref{5A} but exceeds its experimental value $T_{_{\rm C}}^{\rm Exp}=560\,$K. This discrepancy arises because the MFA  does not account for the longitudinal and transverse fluctuations of the spins $\bm S_j$, which influence the given spin  $\bm S_i$. We will revisit this issue in further discussion below.
\subsection{\label{ss:low-T}Spin-wave supression of  the YIG magnetization}  
 \begin{figure} 
 \includegraphics[width=.95 \columnwidth]{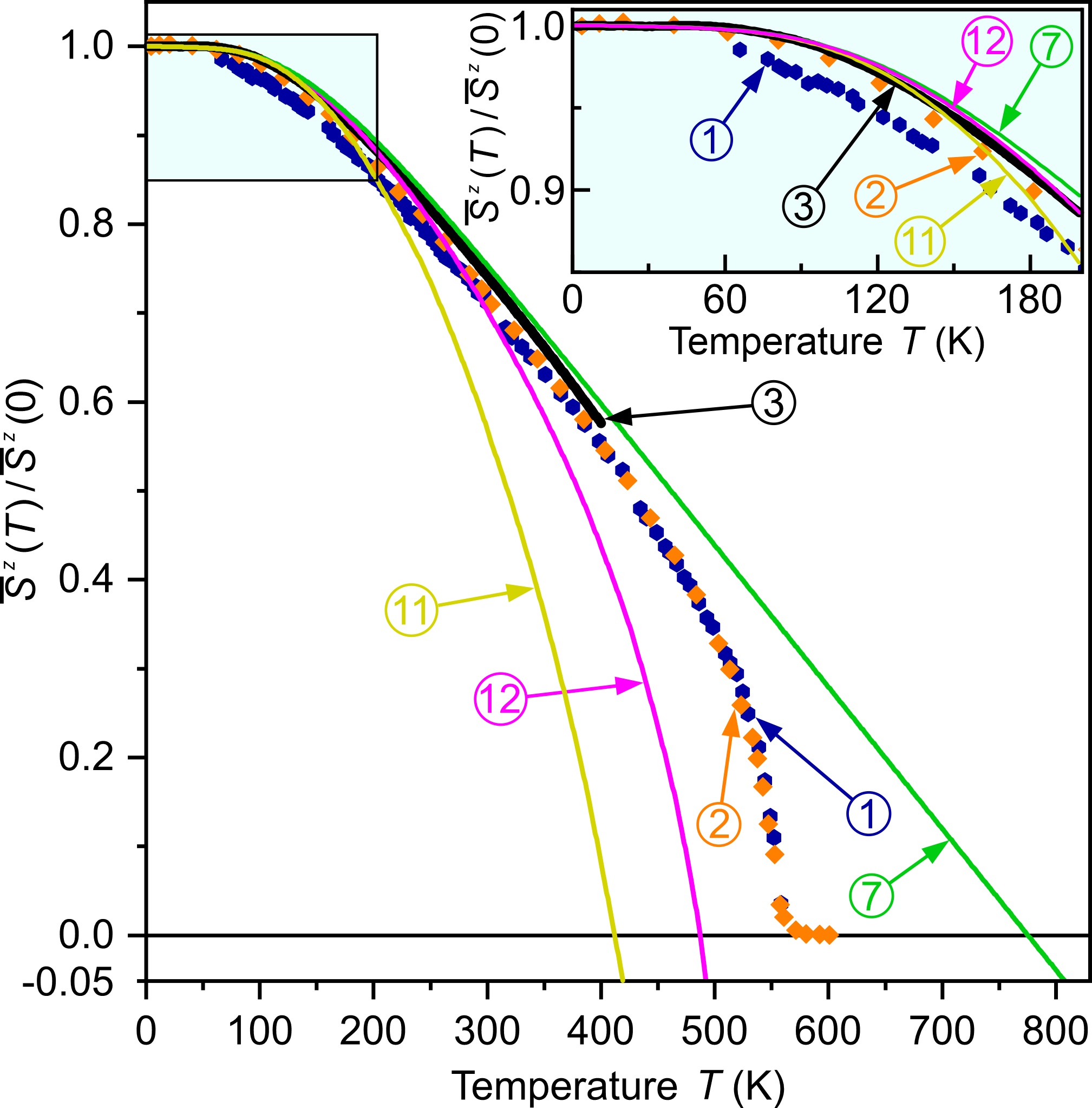}\\  
   \caption{\label{F:5}Experimental and numerical results for $\overline{S}^{z}/\overline{S}^{z}(0)$ vs the temperature $T$.  Experimental results \textcircled{1},  \textcircled{2}, and \textcircled{3}  are the same as in  Fig.\,\ref{F:2}.  \\ 
   Numerical solutions of Eq.\,\eqref{S-Td} are represented by solid lines \textcircled{7}, \textcircled{11}  and \textcircled{12}. For the green line \textcircled{7} (also shown in Fig.~\ref{F:2}) we  used Eq.\,\eqref{S-Td} with temperature-independent frequencies $\omega_j(k)$. For the yellow line \textcircled{11} we applied in Eq.\,\eqref{S-Td}  temperature-dependent frequencies $ \omega_j(k,T)$, Eq.\,\eqref{T-dep}, and for the magenta line \textcircled{12} $\omega_j(k,T)$--Eq.\,\eqref{T-dep1} with $\delta=0.5$. For both, we used the experimental results for magnetization $\overline{S}^{z}$ from the experiment \textcircled{2}.\\
   Inset: high-resolution close-up of the low-$T$ range marked in the main panel as a rectangle with a light-blue background.
   }
\end{figure} 
We demonstrated through Eqs.\,\eqref{smallT} that in the MFA, the deviation of the sublattice magnetization from its $T=0$ values is exponentially small. In contrast, the corrections due to the magnon contribution exhibit only power-like smallness. To determine these corrections, we need to find analytical expressions for spin-wave corrections to the YIG magnetization in the thermodynamic equilibrium.   
As previously mentioned, the magnetization $M(T)$ is proportional to the average (in time) value of the $z$-component of the total spin $\overline{S}^{z}$ in the elementary cell.  By incorporating the Holstein-Primakoff representation\,\eqref{HP} into Eq.\,\eqref{Sa}, we arrive at the conclusion that 
\begin{subequations}
	\begin{align}\begin{split}\label{S-Ta}
			\overline S^{z}(T)=\, &   \Big [  \sum_{j=1}^{n_+}\big ( S - \langle a_{jn}^\dag a_{jn} \rangle \big ) \\
			& - \sum_{j=n_++1}^{n}\big ( S - \langle a_{jn}^\dag a_{jn} \rangle \big ) \Big ] \ . 
	\end{split}\end{align}
	After the Fourier transform, Eq.\,\eqref{S-Ta} becomes
	\begin{align}\begin{split}\label{S-Tb}
			\overline S^{z}(T)= \overline S^{z}(0) - \sum_{\bm k} \Big [  & \sum_{j=1}^{n_+} \langle a_{j\bm k}^\dag a_{j\bm k}\rangle    - \sum_{j=n_++1}^{n} \langle a_{j\bm k}^\dag a_{j\bm k}\rangle   \Big ] \,, 
	\end{split}\end{align}
	where $ \Delta n = n_+ - n_-$is equal to four in YIG.
	Note that the Bogolyubov canonical transformations \eqref{uv}, which transform the Hamiltonian  \eqref{Ham2A} to the diagonal form~\eqref{10}, leave invariant the quadratic form in Eq.\,\eqref{S-Tb}: 
	\begin{align}\begin{split}\label{S-Tc}
			& \sum_{j=1}^{n_+} \langle a_{j\bm k}^\dag a_{j\bm k}\rangle    - \sum_{j=n_++1}^{n} \langle a_{j\bm k}^\dag a_{j\bm k}\rangle \\  = &\sum_{j=1}^{n_+} \langle b_{j\bm k}^\dag b_{j\bm k}\rangle    - \sum_{j=n_++1}^{n} \langle b_{j\bm k}^\dag b_{j\bm k}\rangle\ .
	\end{split}\end{align} 
\end{subequations}  
Then, Eq.\,\eqref{S-Ta} takes on the   form: 
\begin{align} \label{S-Td} 
	\overline S^{z}(T)=   \overline S^{z}(0)  - \sum_{\bm k} \Big [   \sum_{j=1}^{n_+} n_{j+}(\bm k)  - \sum_{j=n_++1}^{n} n_{j-}(\bm k)   \Big ] \ .
\end{align} 

The occupation numbers of magnons are denoted as  $ n_{j+}(\bm k)=\langle b_{j\bm k}^\dag b_{j\bm k}\rangle$  for $n_+$   FM  modes (where  $j = 1, 2, \dots, n_+$ ) and as $ n_{j-}(\bm k)=\langle b_{j\bm k}^\dag b_{j\bm k}\rangle$  for $n_-$   AFM  modes (with   $j = n_++1, \dots, n$ ).

For future reference, it will be helpful to express Eq.\,\eqref{S-Td}  in a more concise form:
\begin{subequations} \label{S-T} \begin{align}  \label{S-Te}
		\overline S^{z}(T)= &  \overline S^{z}(0)   - {\cal N}_{+}  + {\cal N}_{-} \,, \\  \label{S-Tf}
		{\cal N}_{+}= \,& \sum_{j=1}^{n_+} \langle  n_{j+}\rangle_{\bm k} \,,\quad 
		{\cal N}_{-}=   \sum_{j=n_++1}^{n} \langle  n_{j-}\rangle_{\bm k} \,, 
\end{align}\end{subequations}
Here, for any function $f(\bm k)$ the mean value $\langle f (\bm k) \rangle_{\bm k} $ in  the $\bm k$-space over the first Brillouin  zone is defined as follows
\begin{equation}\label{not}
	\sum_{\bm k} f (\bm k) = \langle f (\bm k) \rangle_{\bm k} 
	=\frac v {(2\pi)^3} \int f_{\bm k} d^3 k\ ,
\end{equation}
where $v =  a^3 /2$ is the volume of the elementary cell. The factor  $\frac{1}{2}$ appears because, in BCC crystals, the elementary cell occupies half of the cube with size  $ a$. Noting that $\int d^3 k =  (2\pi)^3/v $, we see that $\langle 1 \rangle_{\bm k} = 1$. 

The Eq.\,\eqref{S-Td} represents a simple and significant conclusion,   announced above:  the excitation of one ferromagnetic magnon (with frequencies  $\omega_{\rm d}(\bm k)$  and  $\omega_{\rm da}(\bm k)$,  represented by green lines in Fig.\,\ref{F:3})  decreases the mean spin of YIG by unity. In contrast, the excitation of one antiferromagnetic magnon (with frequencies  $\omega_{\rm a}(\bm k)$  and  $\omega_{\rm ad}(\bm k)$,  indicated by blue lines in Fig.\,\ref{F:3})  increases the mean spin by unity. To determine the corrections to the magnon-induce corrections to the temperature dependence of YIG magnetization in thermodynamic equilibrium, we need to substitute in \eqref{S-Td} the Bose-Einstein distribution  for their occupation numbers:
\begin{equation}\label{BE}
	n_j(\bm k) = \Big \{  \exp \Big [ \frac {\omega _j (\bm k)}{T}-1 \Big ] \Big \}^{-1} \ , 
\end{equation}  
where $\omega _j (\bm k)$ are the magnon frequency spectra. 

  These spectra have been the focus of numerous studies, starting with a paper by Douglass in 1960 \cite{Douglass1960}, followed by a report by Harris in 1963 \cite{Harris1963}, among others. Cherepanov, Kolokolov, and L'vov conducted a comprehensive analytical study of the 20 magnon branches in YIG, as detailed in their 1993 publication \cite{Cherepanov1993}. In the graphical form, the result of Ref.\,\cite{Cherepanov1993} is shown in Fig.\,\ref{F:3}. However, the full expressions for the magnon frequencies provided in Eq. (6) of \cite{Cherepanov1993} are very cumbersome. For our needs, we hope that using the simplified version detailed in Appendix \ref{E} will be sufficient.
  
 Note that our final conclusion \eqref{S-T} regarding the temperature dependence of $\overline S^z(T)$ is based solely on the total number of magnons  ${\cal N}_\pm$ in each frequency branch. As a result,  the function $\overline S^z(T)$ is very weakly dependent on the specifics of $\omega_j(\bm k)$. This allows us to use either version of $\omega_j(\bm k)$ from Ref. \cite{Cherepanov1993} or from \cite{Princep2017a, Princep2017j} without any noticeable changes in the solution. 

In Fig.\,\ref{F:5}, we plot the temperature dependence of the mean spin  $\overline S^{z}(T)$, represented by the solid green line \textcircled{7}, calculated from Eq.\,\eqref{S-Tb} (see  this line also in Fig.\,\ref{F:2}). This calculation was performed using $T$-independent dispersion relations  $\omega_j(k)$, as explained in Appendix \ref{A6}. Formally, Eq.\,\eqref{S-Tb} is valid in the low-temperature limit, where $  t \ll T_{_{\rm C}} \approx 560 \, \text{K}$. Interestingly, the theoretical dependence closely matches the experimental results of $\overline S^{z}(T)$ up to about  $  T \simeq 400 \, \text{K} $ Recently, a similar result was obtained in \cite{Barker2019} (compare the blue line in their Fig. 2 with our green line \textcircled{7} in Fig.\,\ref{F:5}) through direct numerical simulation of the Landau-Lifshitz equation \,\cite{landau1980} supplemented by a stochastic random force. This force is governed by the fluctuation-dissipation theorem \,\cite{landau1980}, mimicking the effect of the quantum thermostat with the Bose-Einstein distribution \,\eqref{BE}.

Note that all experiments were conducted with the external magnetic field $  H_{\text{ext}} $ which creates an internal magnetic field in the sphere given by $  H_{\text{int}} = H_{\text{ext}} - {4\pi M}/{3} $. This field introduces an additional frequency, $ \omega_0 = \gamma H_{\text{int}}$ for  FM modes, while subtracting this frequency from the  AFM  modes. In terms of temperature, $ \omega_0 \approx 7.4 \, \text{K}$for $ H = 1 \, \text{T}$ This value is relatively low compared to typical $ \omega_j$ frequencies, which are usually of the order of a few hundred Kelvin.
Consequently, the external magnetic field does not significantly alter the resulting $M(T)$ dependence (by no more than four percent). In our semi-quantitative approach, this difference can be safely ignored.

A possible reason for the deviation of the green line  \textcircled{7}   from the experimental observations  \textcircled{1}, \textcircled{2}, and \textcircled{3}  in Fig.\,\ref{F:2} is the temperature dependence of the magnon frequencies $\omega_j(k, T)$, which was not taken into account for the calculation for line  \textcircled{7}. This dependence will be discussed in the next section.
 
\subsection{\label{sss:T-dep}Temperature dependence of the spin-wave spectra}
 All expressions for the frequencies analyzed in Appendix\,\ref{E}  are derived using the Holshtein-Primakoff representation, which is valid for the temperature $T=0$~K.  To apply them 
in the intermediate temperature range, especially near the Curie temperature, one must account for the temperature dependence of all 20 spin-wave spectra branches $\omega_j (\bm k)$. A relatively straightforward and simple approach to solve this problem can be derived from the simplest version of the Green's function splitting, as suggested by Tyablikov\,\cite{Tyablikov1967}.   In this method, all analytical expressions for $\omega_j (\bm k)$ are multiplied by the factor $\overline S^{ z} (T)/ \overline S^{ z} (0)$, which equals unity as $T\to 0$ and vanishes when $T\to T_{ _{\rm C}} $.  
\begin{equation}\label{T-dep}
\omega_j (\bm k) \Rightarrow \omega_j (\bm k, T)= \frac  {\overline S^{z}(T) }{ \overline S^{z}(0)  }\omega_j (\bm k)\ .
\end{equation}

This result can be understood from a physical perspective, at least for small $a\,k \ll 1$ and
relatively small temperatures $T \ll T_{_{\rm  C}}$.   Under these conditions, the main contribution to the reduction of $\overline S^{z}_{\rm }(T)$   comes from the rapid spin waves with $a\,k \sim 1$.   This allows us to average the problem over fast motions,  and to consider slow spin waves with  $a\,k \ll 1$, at least in the limit $T \to  0$~K. Therefore, the classical Holstein-Primakoff representation, which is valid at  $T=0$~K, should be replaced by its temperature-modified version for non-zero temperatures. In this modification, $ S$ is replaced by $ S_z(T) $. When this modified version of the Holstein-Primakoff representation is directly substituted into the exchange Hamiltonian, it reveals that the quadratic terms involving $ a $ and $ a^\dag $ are proportional to $\overline S^z(T) $ multiplied by a linear combination of the exchange integrals. This indicates that the same structure — $ \overline S^z(T)$ multiplied by the exchange integrals — will govern the magnon frequencies in ferromagnets and ferrimagnets, including YIG. For instance, the expression for the frequency $\omega(k)$ in Eq.\eqref{Tc} below becomes $ \omega(k,T) = \overline S^z(T) (J_0 - J_k) $. This result is consistent with Eq.\eqref{T-dep}.

 The yellow solid line  \textcircled{11}   in   Fig.\,\ref{F:5}  illustrates the temperature  dependence of the normalized total mean spin $\overline S^{z}(T)/\overline S^{z}(0)$, Eq. \eqref{S-Td},  using temperature dependent frequencies $\omega_j(k,T)$,  Eq.\,\eqref{T-dep},  obtained  with the experimental dependence  of the mean normalized spin \,\cite{Hansen1974}  [orange dots \textcircled{2}].  It crosses the value $\overline{S}^{z}(T)=0$ at $T_\times \approx 410\,$K, which is below the experimental Curie temperature $T^{\rm Exp}_{_{\rm C}} \approx 560\,$K. In contrast, a solid green line    \textcircled{7} in Figs.\,\ref{F:2} and \ref{F:5},   found with temperature-independent $\omega_j(k)$, yields $T_\times \approx 780\,$K, significantly above $T^{\rm Exp}_{_{\rm C}}$.  Clearly, accounting for the temperature dependence of $\omega_j (\bm k, T)$ is necessary. However, the simple Eq.\,\eqref{T-dep}  overestimates the suppression of magnetization casued by the excitation of spin-waves. 

 To resolve this problem, we consider this phenomenon from a physical perspective. Note that when $\omega(k) \ll T$, the Bose-Einstein distribution can be approximated by the Rayleigh-Jeans  distribution 
 \begin{equation}\label{RJ}
n _j (\bm k)= T / \omega_j(\bm k,T)\ .
 \end{equation}
Accordingly, in our approximations, as $T$ becomes sufficiently large, $n(\bm k,T)$ also becomes very large, and $\overline{S}^{z}(T)$ becomes negative. 

On the other hand, according to Eq.\,\eqref{HPa}, $\langle a^\dag a\rangle \leq S$ holds in thermodynamic equilibrium at any temperature. The divergence of the occupation numbers $n(\bm k, T)=\langle a^\dag a\rangle$ is a property of Bose operators. The observed contradiction arises from the approximate nature of the Holstein-Primakoff representation of spin operators in terms of boson operators. 

 A consistent way to resolve this contradiction is to formulate a theory from the very beginning in terms of spin-operators as it was done in the Belinicher-Lvov diagrammatic technique for spin operators \,\cite{Belinicher1984a}. Unfortunately, this procedure is very complicated, even for simple ferromagnetic materials, as was noted in \cite{Kolokolov2025}. 
 Such a project lies well beyond the scope of this paper. As an initial step in analyzing the temperature dependence of the mean spin in the complex 20-sublattice crystallographic structure of  YIG, we propose a very simple method to adjust the contribution from spin waves. This is achieved by modifying Eq.\,\eqref{T-dep}  using a phenomenological tuning parameter $ 0 \leq \delta \leq 1$:
 \begin{align} \label{T-dep1} 
 \omega_j (\bm k, T)=\,& \frac  {\overline S^{z}(T)\  \omega_j (\bm k) }{\sqrt{ [\overline S^{z}(0) \delta]^2 + (1-\delta^2 )[\overline S^{z}(T)]^2 }  }\ .
\end{align} 
    When $\delta=1$, Eq.\eqref{T-dep1} coincides with Eq.\eqref{T-dep}, resulting in the Curie temperature of approximately $ T_{_{\rm C}}\approx 440 \, K $, as indicated by the yellow line \textcircled{11} in Fig.\,\ref{F:5}. When $\delta=0$, the function$ \omega_j(k, T) $ in Eq.\,\eqref{T-dep1} becomes temperature-independent, yielding a Curie temperature of about $ T_C \approx 810 \, K $, represented by the green line \textcircled{7}. For an intermediate value of $\delta=0.5$, the magnetization $ M(T) $ is illustrated by the magenta solid line \textcircled{12}, which gives $T_{_{\rm C}} \approx 490 \, K $. The experimental Curie temperature $ T_{_{\rm C}}\approx 560 \, K$ corresponds to $\delta=0.428$, with the resulting $ M(T) $ shown as the red solid line \textcircled{8} in Fig.\,\ref{F:2}. This curve closely matches the experimental data for $ M(T) $ presented as blue and orange dots \textcircled{1} and \textcircled{2}, along with the black solid line \textcircled{3} in Fig. \ref{F:2}.

We note that Eq. \eqref{T-dep1} was proposed as an ``educated guess", justified solely by the successful comparison of the resulting $(M(T)$ with experimental data. The analytical derivation of this equation (if possible), along with a direct comparison of the proposed dependence $\omega(\bm k, T)$ with experimental observations, will be addressed in the future.
Additional details about the differences in our approach to EuO and YIG can be found in Appendix \ref{s:comp}.

\section{\label{ss:SW-improvement}  Unifying the spin-wave and mean-field approximations }
As discussed earlier, the mean-field approximation qualitatively describes the behavior of $ \overline{S}^{z}(T)$ over a wide temperature range. It predicts the Curie temperature $ T_{_{\rm C}}$ as the point where $\overline{S}^{z}(T) > 0$ for $ T < T_{_{\rm C}}$ and $ \overline{S}^{z}(T) = 0$ for $ T > T_{_{\rm C}}$ in the absence of an external magnetic field. However, the MFA fails to capture the power-like behavior of $\overline{S}^{z}(T)$ in the low-temperature region, particularly when $ T \ll T_{_{\rm C}}$. 

In contrast, the spin-wave approximation (SWA) in  YIG   correctly describes this dependence, in particular up to $ T \approx T_{_{\rm C}}/2$. On the other hand, SWA fails near $ T_{_{\rm C}}$, predicting negative values for $\overline{S}^{z}(T)$ at temperatures above a certain point, where $ \overline{S}^{z}(T)$ crosses the value  $\overline S^{z}(T)=0$. 

It would be beneficial to unify the  MFA  and the  SWA  to develop an approach that accurately captures the power-like dependence of $ \overline{S}^{z}(T)$ as described by the SWA in the low-temperature range while capturing the physically accurate representation of $ \overline{S}^{z}(T)$ in the vicinity of the critical temperature $T_{_{\rm C}}$, similar to how MFA does on both sides of $ T_{_{\rm C}}$.

In the simplest case of one-sublattice ferromagnets, this program was implemented in Ref.~\cite{Kolokolov2025} and is briefly outlined in the next Section \,\ref{ss:EuO}. We extend this program to multi-sublattice ferrimagnets, such as YIG, in Sec.\,\,\ref{ss:YIG}.


\subsection{\label{ss:EuO} Unified SW  and MFA  model \\  in simple single-sublattice ferromagnets }
The original MFA does not consider the spin-wave correction to the average spin. An effective method for improving this deficiency was discussed in detail in Ref.~\cite{Kolokolov2025}, focusing on the case of ferromagnetic materials  EuO and EuS, which have a single magnetic ion in their elementary cell. In \cite{Kolokolov2025}, Eq.\,\eqref{MFA} in the original MFA was replaced by the 
 \begin{subequations}
 \begin{equation}\label{SW-MFAa}
 \overline{S}^{z} = b_{_{S}}(y)\,, \quad y= \ln \Big ( 1+ \frac 1 {\cal N}\Big ) \ .
 \end{equation}
Here ${\cal N}=\sum_{\bm k} n_{\bm k}= \langle n_{\bm k} \rangle   $ represents the mean number of magnons in the first  Brillouin zone of a single ferromagnetic branch. In the limit of low temperatures, when ${\cal N}\to 0$, the argument $y\gg 1$. In this case
 \begin{equation}\label{SW-MFAb} 
 \overline{S}^{z}= b_{_{S}}(y)\approx S  - \frac{{\cal N}}{1+ {\cal N}} \approx S - {\cal N} \,,
 \end{equation} \end{subequations}
i.e. as anticipated: the decrease in $\overline{S}^{z}$ is  governed by the exitation of spin waves\,\cite{Bloch1930}. 

The equation for the Curie temperature, as suggested   in \cite{Kolokolov2025}, generalizes Eq.\,\eqref{3} by presenting it in terms of the spin-wave frequency $\omega_k$ in EuO:
\begin{align}\begin{split}\label{Tc}
T_{_{\rm C}}^{^{\rm MF-SW}}=\,& \frac{S(S+1)}{3}\Big \langle \frac1{J_0-J_k}\Big \rangle ^{-1}\\ 
= \,&\frac{S+1}3 \Big \langle \frac1{\omega_k}\Big \rangle ^{-1} \,, \quad \omega_k = S(J_0 -J_k) \ .
\end{split}\end{align} 
\subsection{\label{ss:YIG} Unified SW-MFA minimal model in multi-sublattice ferrimagnets} 
The goal of this section is to suggest the simplest model for unifying MFA with SWA in multi-sublattice collinear ferrimagnets, such  as   YIG,  while still 
accounting for the fundamental physics involved.

 In semi-qualitative analysis, we accepted the approximation  Eq.\,\eqref{T-dep}, where $\omega_j(\bm k, T)$ is independent of $\overline{S}_{\rm a}^{z}  $ and $\overline S_{\rm d}^{z}$ individually, being proportional to the total spin in the elementary cell $\overline{S}^{z}$.
In this case, it is superfluous to analyze two equations for $\overline{S}_{\rm a}^{z} $ and $\overline S_{\rm d}^{z}$, and we only need to consider one equation for $\overline{S}^{z}$ (equal to $\Delta n\, S$ for $T=0$ with $\Delta n=4$ in YIG).
Within this framework, and using  Eqs.\,\eqref{S-T},   we propose the following  revision of   the original  MFA Eqs.\,\eqref{4}:
\begin{align}  \label{26}
\overline S^{z}=\,&  \,  b _{_{\Delta n\,S }}(y)\,, \quad  y= \ln \Big [ 1+ \frac 1 { ({\cal N}_{+}- {\cal N}_{-})}\Big ]\ .   \end{align}
Here,  according to 
Eq.\,\eqref{S-Tf}, ${\cal N}_{+}$ and ${\cal N}_{-}$ represent the mean values of the magnon numbers in the $\bm k$-space over the first Brillouin zone, for $n_+$ ferromagnetic branches and $n_-$ antiferromagnetic branches, respectively.

 This equation coincides with a single-sublattice version of the Unified SW and MFA model \eqref{SW-MFAa} if one replaces $\Delta n$ with unity and ${\cal N}_+-{\cal N}_-$ with ${\cal N}$.

In  the low-temperature range where ${\cal N}_{\rm a} \ll 1$,  ${\cal N}_{\rm d} \ll 1$ and $1/{\cal N}_{\rm a} \gg 1$,  $1/{\cal N}_{\rm d} \gg 1$ we, according to Eq.\,\eqref{26} and  expansion\,\eqref{expA},  
  can ultimately arrive at the desired conclusion:
 \begin{align}\begin{split}\label{SA}
   \overline S^{z}(T) =\,\,&   \Delta n\,S  -   {\cal N}_{+}+   {\cal N}_{-}  
   = \Delta n\,S \\ &-  \sum _{j=1}^{n_+}\langle n_{j+}(\bm k)\rangle _{\bm k} 
   +\sum _{j=n_++1}^{n}\langle n_{j-}(\bm k)\rangle _{\bm k}  \ .
 \end{split}\end{align} 
This is the exact relation given by \eqref{S-Tb} for the spin-wave correction to the mean spin in YIG, which was not included in the original  MFA. As discussed earlier, the dependence of $ \overline{S}^{z}(T)$, shown in Fig.\,\ref{F:5} by solid green line \textcircled{7}, aligns well with experimental observations in YIG from $ T=0$ up to approximately $ 400\,K$.

Our next objective is to derive a new expression for the Curie temperature  $T_{_{\rm C}}^{^{\rm MF-SW}}$  in multi-sublattice ferrimagnets, such as YIG. At this temperature, the average spin $  \overline{S}^{z}(T) $ approaches zero as the temperature $  T $ approaches the Curie temperature from below. Note that 
as $\overline S^{z} \to 0$, and, consequently, according to \eqref{T-dep}, $\omega _j(\bm k, T)\to 0$, 
the occupation numbers $\langle n_j(k) \rangle   \to \infty$.
  In the considered case, the Bose-Einstein distribution \eqref{BE} can be approximated by the Rayleigh-Jeans distribution \eqref{RJ} in which we have to account for temperature dependence of $\omega(k,T)$ according to Eq.\,\eqref{T-dep1}: 
\begin{align} \begin{split}\label{RJ1}
 n_{j\pm} (\bm k, T)=\,& \frac{T}{\omega_{j\pm}(\bm k, T)}  \ .
 \end{split} \end{align} 
We denote the frequencies in  FM  (d-sites in YIG) and  AFM  (a-sites in YIG) modes as $\omega_{j+}(\bm k)$ and $\omega_{j-}(\bm k)$, respectively.
Using only the first term in the expansion \eqref{expB},
we derive from Eqs.\,\eqref{26} and \eqref{RJ} a homogeneous linear equation for $\overline S^{z}$ that has a non-zero solution below a certain temperature $T_{_{\rm C}}^{^{\rm MF-SW}}$.  This solution approches zero as $T\to T_{_{\rm C}}^{^{\rm MF-SW}}$. In this solution  
\begin{align} \begin{split}\label{CT2}
   T_{_{\rm C}}^{^{\rm MF-SW}} =    \frac{\Delta n S +1}{3\, \delta} & \Big [    \sum_{j= 1}^{n_+}\Big \langle \frac 1{ \omega _{j+}(\bm k)} \Big \rangle \\   - &   \sum_{j= n_++1}^{n}\Big \langle \frac 1{ \omega _{j-}(\bm k)} \Big \rangle \Big] ^{-1}  .
  \end{split} \end{align}  
 Recall, for   YIG,  $n_+=12$,  $n_-=8$  and $\Delta n \equiv n_+ - n_- =4$. Using the magnon frequencies from \cite{Princep2017a,Princep2017j} (cf. the right part of Tab.\,\ref{t:2}), we fix the 
 tuning parameter $\delta$ in  \eqref{CT2} such that $T_{_{\rm C}}^{^{\rm MF-SW}}$ agrees with the experimental value of the Curie temperature. For $\delta = 0.428$,  $T_{_{\rm C}}^{^{\rm MF-SW}}\approx 560\,$K in full agreement with the experiment. Recall that  $ \delta=0$ in \eqref{T-dep1} corresponds to the temperature-independent $ \omega_j(\bm k)$, while for $ \delta=1$,  $\omega_j (\bm k,T )$ is strongly dependent on $ T$ according to \eqref{T-dep}. 

To determine the temperature dependence of $ \overline S^{z}(T)$ in the unified SW-MFA minimal model  \eqref{26}, we numerically solved this equation using $ {\cal N_{\rm d}}$ and $ {\cal N_{\rm a}}$ from Eqs.\,\eqref{S-Td} and \eqref{BE}, along with temperature-dependent frequencies from Eq. \eqref{T-dep1} with   $ \delta = 0.428$.  For more details on the solution procedure, see Appendix\,\ref{A6}.  The result, shown in Fig.\,\ref{F:2} as the red solid line \textcircled{8}, was compared to the experimental data represented by \textcircled{1}, \textcircled{2}, and \textcircled{3}.
Taking into account that the simple model Eq.\,\eqref{26}  has only one tuning parameter $\delta$, we find that the quantitative agreement between the theory and experiment across the entire temperature range from $T = 0$ up to the Curie temperature $T_{_{\rm C}}^{^{\rm MF-SW}}$ is more than satisfactory. It suggests that the minimal model \eqref{26} adequately captures the basic physics of the problem.

\section{\label{s:sum} Discussion and Summary}

 In this paper, we suggested a theoretical description of the temperature dependence of spontaneous magnetization $\overline S^{z}(T) $ in multi-sublattice collinear ferrimagnetic 
 materials in the entire temperature range from $T=0$ to the Curie temperature $T_{_{\rm C}}$. As an example of such a material, we have chosen the  Yttrium Iron Garnet 
 with a 20-sublattice magnetic structure.
We combined two well-known approaches:  Weiss-Heizenberg mean-field approximation that is accurate in the near-$T_{_{\rm C}}$ range, and spin-wave approximation, describing the suppression of the magnetization due to spin-wave excitations and exact at low-$T$ range. 

We generalized both approaches to the multi-sublattice collinear ferrimagnetic. In the MFA framework, we accounted for two magnetic sub-lattices and derived coupled equations Eq.\,\eqref{4} for the $T$-\,dependence of the mean spins for each sublattice, and the analytical expression in Eq.\,\eqref{5A} for the Curie temperature $T_{_{\rm C}}^{^{\rm MF}}$. 

Within SWA, we accounted for multiple magnon frequency spectra branches and derived Eq.\,\eqref{S-Tb}, which describes the suppression of $ \overline{S}^{z}(T)$ by spin-wave excitations. Given that the exact derivation of the temperature dependence of the magnon frequencies via the diagrammatic technique is too cumbersome for the level of the approximations that we used in this work, we suggested a phenomenological description of $\omega_j(\bm k, T)$, Eq.\,\eqref{T-dep1}, that employs one tuning parameter $\delta$.

The central theoretical part of our paper focuses on the unifying SW and MF approximations for the multi-sublattice ferrimagnets. This approach culminates in Eq.\,\eqref{26} which connects    $\overline{S}^{z}(T)$ with the total amount of magnons in the ferromagnetic and antiferromagnetic branches.
Formally speaking,  Eq.\,\eqref{26} is a  relatively simple generalization of unified SW and MFA
Eq.\,\eqref{SW-MFAa} for the simple one-sublattice ferromagnetics such as EuO or EuS analyzed in Ref.\,\cite{Kolokolov2025}. In Eq.\,\eqref{26}, we replaced the spin of a single ion $  S $ in EuO with the total spin $  \overline{S}^{z}(0) $ in the elementary cell of a multi-sublattice ferrimagnetic. Additionally, we substituted the total number of magnons $  \mathcal{N} $ in a single  FM  branch with the value $ \mathcal{N}_+ - \mathcal{N} _- $, which accounts for magnons from all  FM and  AFM branches in the multi-sublattice ferrimagnets.
These replacements are not justified by some consistent procedure that is expected to lead to much more complicated basic equations. A price for simplicity of the current version of a theory for magnetization in multi-sublattice ferrimagnets is the tuning parameter $\delta$ in Eq.\,\eqref{T-dep1} for $\omega_j(\bm k,T)$, which was chosen to require the theoretical Curie temperature to match its experimental value. 

The unified SW-MFA minimal model with the chosen tuning parameter $\delta=0.428$  demonstrate good quantitative agreement between the theoretical $\overline{S}^{z}(T)$ and experimental magnetization temperature dependencies of YIG in the entire temperature range from $T=0$\,K to $T=T^{\rm Exp}_{\rm _C}$. 

Within MFA, it is expected that the magnetization $\overline{S}^{z}(T)$ scales as $\sqrt{T_{_{\rm C}}^{^{\rm MF}}-T}$ near the Curie temperature.  
Curiously, we have found that in YIG, all the experimental  $\overline{S}^{z}(T)$  shown in Fig.\,\ref{F:2} as \textcircled{1}, \textcircled{2}, and  \textcircled{3}, and the theoretical prediction of the SW-MFA minimal model \textcircled{8}, are fit well by $\overline{S}^{z}(T) \propto (T_{_{\rm C}} - T)^\beta$ with the scaling exponent $\beta\approx 0.50 \pm 0.005$ in almost entire range of $\overline{S}^{z}(T)/\overline{S}^{z}(0)$ as it decreases from about 0.95 to zero.

We thus definitely demonstrated that $  M(T) $ scales in  YIG  nearly normally as $  \sqrt{T_{\rm _C} - T} $ for all temperatures. This behavior contrasts sharply with the anomalous scaling of $M(T)$ in simple ferromagnets EuO and EuS, where $\beta \approx 1/3$\,\cite{Kolokolov2025}. This fact highlights the need for further investigation of the magnetization of different materials.

 Our theoretical approach marks a significant advancement in the description of magnetic systems. We demonstrated the effectiveness of unifying two simple theoretical approximations for the description of complex magnetic materials.
We believe that the applicability of our SW-MFA minimal model is not limited to YIG but may be extended to various multi-sublattice ferrimagnets and antiferromagnets. A further generalization of our minimal model will allow us to study more complex magnetic structures, such as Gadolinium Gallium Garnet (GGG, Gd$_3$Ga$_5$O$_{12}$) at low temperatures \cite{Petrenko1998,Deen2015,Oyanagi2019,Serha2024}. GGG has the same crystal structure and lattice parameters as  YIG, making it an ideal substrate material for the epitaxial films of YIG. These layers have recently been used in research on magnon lifetime and transport at millikelvin temperatures \cite{Knauer2023, Serha2025,Schmoll2025_2,Schmoll2025}, paving the way for advancements in quantum magnonics.

  \section*{Acknowledgments}
This research was funded in part by the Austrian Science Fund (FWF) project Paramagnonics [10.55776/I6568]. 

\appendix
\section{\label{A:dem}Primary reasons for low magnon damping  in YIG}
YIG is known for its exceptionally narrow linewidth of spin-wave resonance due to the low damping of excited magnons. There are three main reasons for this\,\cite{Cherepanov1993}. 

\textbullet\  Firstly, there is no magnon-phonon coupling.
 Fe$^{3+}$ ions have a half-filled orbital, which means that the orbital angular momentum $L = 0$. As a result, the spin-orbit interaction, which is proportional to $\bm{S} \cdot \bm{L}$ and is responsible for magnon-phonon coupling,  is also suppressed. Consequently, the magnon damping, which occurs as a result of their decay into phonons and magnons, is also eliminated.
 
\textbullet\ Secondly, YIG has no free electrons and thus no damping mechanisms related to them. 

\textbullet\ And last, but not least, the LPE technique for YIG films\,\footnote{Liquid phase epitaxy is a crystal growth technique used to deposit thin, single-crystal layers of materials onto a substrate. It involves growing a layer from a liquid solution or melt that is supersaturated with the desired material. The substrate is immersed in the solution, and as the temperature decreases, the material precipitates out of the solution and onto the substrate, forming an epitaxial layer  }   and the Czochralski method for bulk YIG  allow this material to be grown with a high level of crystalline perfection, and with a low density of impurities and defects \cite{Tolksdorf1968,Tolksdorf1978,Glass1988,Dubs2020}. 
 As a result, phonon damping in  YIG  is lower than in quartz monocrystals,   particularly at lower temperatures and higher frequencies.
 Similarly, the magnon damping caused by their scattering on impurities and defects is also minimal.
 
The only relevant mechanisms of magnon damping arise from magnon-magnon interactions, which decrease with decreasing temperature.  In a recent study, it is demonstrated that, by effectively draining the magnon thermal bath, magnon lifetimes of up to 18\,µs can be achieved in the highest-purity YIG samples~\cite{Serha2025_2}.

\section{\label{s:comp} Comparison of theoretical approaches to $M(T)$   dependence   in single-    and multi-sublattice magnetics }

In this Appendix, we clarify the similarities and differences between our approach to EuO and EuS ferromagnetics, as presented in our previous paper\,\cite{Kolokolov2025}, and our discussion of YIG developed in the current work.

EuO and EuS are the simplest objects available for theoretical description: they are ferromagnetic materials with only one magnetic ion in the elementary cell and have the smallest applicability parameter $1/Z$  for the mean field approximation (MFA)  with $ Z \approx 12$. These two aspects allowed us to develop a consistent step-by-step analysis of the temperature dependence of spontaneous magnetization $M(T)$, including what is referred to in this paper as “SW-MFA” minimal model for the single-sublattice ferromagnet.   In contrast to YIG, this approach describes the experimental data in EuO and EuS quite well. The main reason for this success is the small value of the applicability parameter $1/Z\approx 1/12$.

Additionally, the simple crystallographic structures of EuO and EuS allow us to develop a suitable diagrammatic approach to calculate the $1/Z$ corrections to SW-MFA. As a result, we obtained excellent quantitative agreement between our theoretical predictions (which did not involve any fitting parameters) and experimental data, as noted in Ref.\,\cite{Kolokolov2025}.
 
Unfortunately, at this stage, we cannot fully apply this method in the study of YIG. The absence of a small applicability parameter for the SW-MFA approach in YIG (where $Z \approx 5$) leads to only a qualitative agreement between the SW-MFA approach and experimental results. Additionally, the complexity of YIG's 20-sublattice crystallographic structure makes the analytical calculation of $1/Z$ corrections difficult and even impractical, as $1/Z$ is not small. In view of these challenges, we limited our study of YIG at the SW-MFA level, achieving quantitative agreement with experiments by introducing a tuning parameter $\delta$.

\section{\label{ss:A1} Positions of Fe$^{3+}$ ions in elementary cell of YIG}

To define the number of nearest neighbors and the relevant distances to assign the coordination spheres, we have drawn in Fig.\,\ref{f:1}\,(b) the Fe$^{3+}$ ions, oriented such that (a)-Fe$^{3+}$ ions are placed in the corners and the center of the half-cube.
  The particular Fe$^{3+}$ ion located at (a)-position (0,0,0)  (see coordinate labels in Fig.\,\ref{f:1}\,(a)) has  $Z_{\rm aa} =8$ nearest-neighbor  (a)-Fe$^{3+}$ ions in positions $(\pm \frac 14, \pm \frac 14,\pm \frac 14)$, separated by distance $\Delta_{\rm aa}= \dfrac{\sqrt 3}{4}\approx 0.433$.  The coordinates of the nearest  $Z_{\rm ad} =6$ neighboring sites (d) with respect to the (a)-Fe$^{3+}$ ion at $(0,0,0)$ are ($0,\frac14,-\frac18$),\ \ ($0,-\frac14,\frac18$); \
($-\frac18,0,-\frac14  $),\ \ ($\frac18,0,\frac14$);
($\frac14$,     $-\frac18$, 0),   ($-\frac14,\frac18,0$), 
  with the separation $\Delta_{\rm ad}= \dfrac{\sqrt 5}{8}\approx    0.2795$.    
  
To find the environment of the (d)-Fe$^{3+}$ ion, we place it at the origin. Then, there are $Z_{\rm dd}=4$ nearest (d)-neighbors with the coordinates  ($-\frac18,     \frac18,-\frac14$),   ($-\frac14,\frac18,-\frac18$),($-\frac14,-\frac18,\frac18$) and ($\frac18,\frac18,\frac14$) at  the distance $\Delta_{\rm dd}=\dfrac{\sqrt 6}{8}\approx 0.306$.\,\,This central (d)-Fe$^{3+}$ ion has $Z_{\rm da}=4$ (a)-Fe$^{3+}$ neighbors at positions ($0, -\frac18, -\frac14$),(0,$-\frac18,     \frac14$),($\frac14,\frac18,0$), (-$\frac14,\frac18,0$) with the separation  $\Delta_{\rm da}=\Delta_{\rm ad}=\dfrac{\sqrt 5}{8}\approx    0.2795$.

\section{\label{ss:J} Exchange integrals  }
\begin{table*}
\begin{tabular}{c | rc c|ccc| c | c| c}
 \# & Reference & Year & Method& $J_{\rm ad}$ & $J_{\rm aa1} \ | \ J_{\rm aa2}$  & $J_{\rm dd}$ & $T _{_{\rm C}}^{^{\rm MF}}$ &  $  \overline S_{\rm d}^{z} /\overline S_{\rm a}^{z} $ & $\omega_{\mathrm{ex}}$ \\ \hline 
 1&  Harris\,\cite{Harris1963} & 1963 & Spin-wave (SW) magnetization& -(92--117) & -(18--23) & -(18--47) & 992--1122 & -$0.82
$& 80 \\ 

 2&  Plant\,\cite{Plant1977} & 1977 & Neutron scattering of SW & -79.6;-74.4 & -16.0;0.0 & -16.0;-10.4 & 836;1004
 & -$0.82$  & 66;107 \\
 
 3&      Cherepanov\,\cite{Cherepanov1993} & 1993& Neutron scattering of SW &  -79.6 & -7.6 & -26.8 & 894 & -0.82 & 80 \\ 
 4&         Oitmaa\,\cite{Oitmaa2009} & 2009& Educated guess, based on \cite{Cherepanov1993} &  -79.6 & -7.6 & -26.8 & 894 & -0.82 & 80 \\ 
 5& Xie\,\cite{Xie2017} & 2017 & First principle + $T _{_{\rm C}}  ^{\rm Exp}$  fitting  & -73.8;-60.0  & -2.6;-2.2 & -5.2;-3.8 & 992;810
 & -0.82  & 85--104 \\
   6&       Princep\,\cite{Princep2017a} & 2017 & Improved Neutron scattering of SW &    -79   & ~~ 0.0   $|$ -12.8~ & -6.0 & 1018 & -0.82& 86 \\
 7&       Gorbatov\,\cite{Gorbatov2021} & 2021 & Fully realistic first principle &    -74.2    &    -0.94 & -5.2 & 1020 & -0.82& 109 \\
\end{tabular}  
\caption{\label{t:1}  Imporant parameters of YIG. The columns from left to right: the reference to the data origin and the method used to extract it; the exchange integrals $J_{ij}$ (values in K),  the Curie temperature  $T _{_{\rm C}}^{^{\rm MF}} $ in the mean-field approximation computed using Eq.\,\eqref{5A}; the ratio  $  \overline S_{\rm d}^{z} /\overline S_{\rm a}^{z} $ for $T \to  T _{_{\rm C}}^{^{\rm MF}}$ from below,  Eq.\,\eqref{5B}; the exchange frequency $\omega_{\rm ex}$ (in K)  according to \eqref{frec1}. Estimates of  $\omega_{\rm ex}$  obtained by various authors, using a number of techniques, vary from 40 to 100\,K,  see Tab.\,1 in \cite{Srivastava1987}. The experimental value of YIG Curie temperature is $T _{_{\rm C}}^{\rm Exp}\approx 559\,$K. For the comparison with other published data, note that $1\,$meV $\approx 11.6\,$K. } 
\end{table*}

\begin{table*}
\begin{tabular}{|| c c | ccccc || ccccc || } \hline \hline 
&  &Saga &of &YIG,& Ref.\,\cite{Cherepanov1993}&\& Fig.\,\ref{F:2}&Final&Chapter&in the &Saga,& \cite{Princep2017a,Princep2017j}  \\  \hline  &&&&&&&&&&& \\
$j$ & mode  & $\omega_0$ & $ \omega _{\rm ex}$&   $\omega_{\rm max}$ &   $ \Big \langle \dfrac 1 {\omega(k)} \Big \rangle $& $ \Big \langle \dfrac 1 {\omega(k)}   \Big \rangle^{-1} $  &   $\omega_0$ & $ \omega _{\rm ex}$ &   $\omega_{\rm max}$ &   $ \Big \langle \dfrac 1 {\omega(k)} \Big \rangle $& $ \Big \langle \dfrac 1 {\omega(k)}   \Big\rangle^{-1}$ \\ \hline 
1 & da1 & 0 & --& 716& 0.00246& 407 &0 & --& 716& 0.00246& 407 \\
2 & da4 & 260 & 3.3& 340& 0.00326& 306 &405&1&429&0.00238&419\\
3 & da9-1 & 290 & 3.4& 372& 0.00296& 338 &405&2.8&472&0.00225&444 \\
4 & da9-2 & 290 & 4.7& 403& 0.00281& 355  &405&2.8&472&0.00225&444\\
5 & da9-3 & 290 & 4.7& 403& 0.00281& 355  &405&2.8&472&0.00225&444\\
6 & da5-1 & 395 & 0& 395& 0.00253& 395  &545&0&545&0.00183&545\\
7 & da5-2 & 395 & 0& 395& 0.00253& 395  &545&1.8&588&0.00175&571\\
8 & da8-1 & 530 & -2& 481& 0.00199& 500  &600&-3.5&515&0.00182&548\\
9 & da8-2 & 530 & -2& 481& 0.00199& 500  &600&-3.5&515&0.00182&548\\
10 & da8-3 & 530 & -3.1& 455& 0.00206& 484  &600&1&624&0.00162&614\\
11 & da6-1 & 660 & -4& 563& 0.00166& 601  &630&-0.5&618&0.00161&623\\
12 & da6-2 & 660 & -4.2& 558& 0.00167& 598  &685&-2&637&0.00152&656\\ \hline 
13 & ad1 & 400 & --& 1060& 0.00114& 903  &400 & --& 1060& 0.00114& 903\\
14 & ad9-1 & 800 & --& 813& 0.00121& 825  &885&4.2&986&0.00106&945\\
15 & ad9-2 & 800 & --& 813& 0.00121& 825  &885&4.2&986&0.00106&945\\
16 & ad9-3 & 800 & 0& 800& 0.00125& 800  &885&0&885&0.00113&885\\
17 & a-3 & 895 & -1.1& 868& 0.00114& 879  &1090&-2.4&1032&0.00095&1055\\
18 & a7-1 & 1045 & -6.1& 898& 0.00105& 955  &1240&-2.5&1179&0.00083&1204\\
19 & a7-2 & 1045 & -2& 996& 0.00098& 1016  &1240&-2.5&1179&0.00083&1204\\
20 & a7-3 & 1045 & -2& 996& 0.00098& 1016  &1240&-9&1022&0.00090&1107\\ \hline \hline 
\end{tabular}
\caption{\label{t:2} Characteristic frequencies (in K) of 12 ferromagnetic modes (with $j=1,2\dots 12$) and 8 antiferromagnetic modes (with $j=13, 14\dots 20$). The left columns are taken from the frequencies presented in Ref.~\cite{Cherepanov1993}, while the right columns are from \cite{Princep2017a,Princep2017j}.}
\end{table*}

The exchange interactions that govern the magnetic order in YIG are the strongest among magnetic interactions of various natures. The energy of this interaction decreases rapidly with the increase in the distance between magnetic ions. The strongest exchange interaction denoted as  $J_{\rm ad} $ occurs between nearest neighbors: Fe$^{3+}$ ions at the (a) and (d) sites, as shown in Fig.\,\ref{f:1}\,(b) by the blue line denoted (ad). The interactions within (a)- and (d)-sublattices, denoted as $J_{\rm aa}$ and $J_{\rm dd}$, respectively, are relatively weaker. 

Over several decades of studies, there have been numerous suggested exchange integrals for YIG, as shown in Table\,\ref{t:1}.

The suggested values for $J_{\rm ad}$  range from approximately  60\,K~\cite{Xie2017} to about  120\,K~\cite{Harris1963}. Additionally, the values for $|J_{\rm aa}|$ and $|J_{\rm dd}|$ primarily fall below 27\,K. The significant variation in the resulting values of exchange integrals can be attributed to the challenges in analyzing different types of experimental data, such as specific heat, paramagnetic susceptibilities, Brillouin scattering, and X-ray scattering. For further details, refer to Ref.\,\cite{Srivastava1982} and Table 1 therein.
Given the significant variation of the suggested values of the exchange integrals that are required for our semi-quantitative analysis of experimental results on the magnetization of YIG, we have chosen to use the values of $J_{(\dots)}$ suggested in Ref.~\cite{Cherepanov1993}. These values fall approximately in the middle of the range of the reported results.

 It was recently observed\,\cite{Princep2017a} that there are two distinct nearest-neighbor bonds (aa), referred to as (aa1) and (aa2), marked in Fig.\,\ref{f:1}\,(b). They have the same lengths but different symmetries. The (aa2) exchange runs exactly along the body diagonal of the crystal and therefore has limited symmetry-allowed components due to the high symmetry of the bond (point group D$_3$). However, the (aa1) exchange, which connects the same atoms with the same radial separation, represents a different Fe-O-Fe exchange pathway due to a different symmetry of the point group (point group C$_2$), so it is different from (aa2) due to the environment around the Fe atoms. In general, $J_{\rm aa1} \neq J_{\rm aa2}$. Precise analysis of neutron scattering data gives  $J_{\rm aa1} \approx 0$, $J_{\rm aa2} \approx -12.8$\,K, and $J_{\rm dd} \approx -6$\,K instead of the previously used value of $J_{\rm dd} \approx -26.8$\,K~\cite{Cherepanov1993}. The exchange interactions between further neighbors, such as $J_{\rm ad1}$ (yellow line), $J_{\rm dd1}$ (light green dd1 line) and $J_{\rm aa3}$ (dark green aa3 line) shown in Fig.\,\ref{f:1}\,(b), are negligible~\cite{Princep2017a}.

We have verified that the suggestion of \cite{Princep2017a}  regarding the inequality of the (ad)-exchange integrals does not significantly affect our qualitative analysis and only complicates the situation. Therefore, we choose the values suggested in \cite{Cherepanov1993}, as listed on line 3 of Tab.\,\ref{t:1} and Eq.\,\eqref{int}.

\section{\label{D} MFA with  dipole-dipole interaction }
To characterize the effect of dipole-dipole interaction on the magnetization of magnetically ordered dielectrics in the MFA, we introduce a characteristic temperature 
\begin{equation}\label{Tdd} 
 T_{\rm dd}\equiv \frac{\mu _{_{\rm B}}M_*}{k _{_{\rm B}}}= 0.067\, \mbox K\ .
 \end{equation}
 Here, $\mu _{_{\rm B}}=9.274\times 10^{-21}$erg/Oe is the Bohr magneton and the Boltsman constant, denoted as  $k _{_{\rm B}}=1.381\times 10^{-16}\,$erg/K. $M_*$ is the typical value of the magnetization, taken for the estimate in the right-hand side of \eqref{Tdd} equal to  $M_*=1000$\,Oe.
 
For example, in the  Weiss MFA, which accounts only for dipole-dipole interaction of classical spins\,\cite{Weiss1907,Kolokolov2025}, the Curie temperature
  \begin{equation}\label{W-MFA}
  T_{_{\rm C}}^{^{\rm W}}=4\pi N_z \mu _{_{\rm B}} M(0)= \frac{4\pi N_z}3 \frac{M(0)}{M_*}T_{\rm dd} \ .
  \end{equation}
 Here $M(T)$ denotes the magnetization. At $T=0$, the magnetization in EuO is   $M(0) \approx 1920\,$Oe, while in YIG, it is about 200\,Oe. The demagnetization factor $N_z\leq 1$ depends on the sample's shape. For spherical samples, $N_z=1/3$. see e.g. \cite{Lvov2012}.    
  
In the presence of  magnetic dipole-dipole interactions, the effective exchange magnetic field  gains an additional contribution known as the demagnetization magnetic field:
\begin{equation} 
H_{\rm dem} = -4 \pi N_z M(T) \ .
\end{equation}

The dipole-dipole interactions in single-sublattice ferromagnets, such as Eu0 and EuS, renormalize the effective magnetic field $H^{\rm eff}$ given by \eqref{heff} as follows 
\begin{align}\label{Heff1}
H^{\rm eff} \Rightarrow \widetilde H^{\rm eff}= H^{\rm eff} -H_{\rm dem} \mu_{_{\rm B}} \ . 
 \end{align}
Appearance of $\mu_{_{\rm B}}$ follows from the fact that $H^{\rm eff} $, traditionally called effective (magnetic) field, actually is the energy of spin in some magnetic field.

Taking into account  Eq.\,\eqref{heff}  along with \eqref{Snorm} and
$M(0)= \mu_{_{\rm B}}S$ one concludes
\begin{align}\label{Heff2} 
 \widetilde H^{\rm eff} = \overline S^z \widetilde {\cal J}_0\,,  \quad
 \widetilde {\cal J}_0= {\cal J}_0- 4\pi N_z \mu^2 _{_{\rm B}   }\ .
\end{align} 
 Next,  using Eq.\,\eqref{3}, one finds the  equation for the correction to the Curie temperature caused by the dipole-dipole interactions: 
 \begin{align}\begin{split}\label{D4} \Delta T = &\frac{4\pi N_z S (S+1)} 3
  \mu^2 _{_{B}}  = \frac{4\pi N_z(S+1)}{3  }  M(0) \mu _{_{\rm B}} \\
  = &\frac{4\pi N_z(S+1)}{3  } \frac{ M(0)}{M_*}  T_{\rm dd} \ .
 \end{split}\end{align} 
An accurate calculation of the correction $\Delta T$ in multi-sublattice ferrimagnets gives the result \eqref{D4} in which $S$ is the total spin of the elementary cell (equal to 10 in YIG) with an additional numerical factor close to unity.

Taking for EuO $M(0)\approx 1920\,$Oe and $S=7/2$, $N_z=1/3$ (for qubic sample), we find
\begin{equation}\label{EuO}\Delta T\approx 2\pi \,T_{\rm dd}\frac{ M(0)}{M_*}\approx 0.79\, \mbox{K} \ .
\end{equation}
A similar estimate for YIG with $M(0)\approx 200\,$Oe and $S=10$ gives
\begin{equation}\label{YIG}\Delta T\approx \frac{44 \pi }{9} T_{\rm dd}\frac{ M(0)}{M_*}\approx 0.21\, \mbox{K} \,,
\end{equation}
which is negligible compared to $T_{_{\rm C}}\approx 560\,$K.

{Analyzing the temperature behavior of $\overline S^z(T)$ with the $\widetilde H^{\rm eff}$, given by \eqref{Heff1} we conclude that the relative contribution of the dipole-dipole interaction is about $T \Delta T/ T_{_{\rm C}}^2 \ll 1$. Therefore, except for the small temperature range around $T_{_{\rm C}}$, in which the critical behavior is observed (as considered e.g. in \cite{Aharony1973,Fisher1973}), we can safely neglect the contribution of the dipole-dipole interaction to the MF behavior of YIG across the entire temperature region.   
  \section{\label{E}Approximations for magnon frequencies $\omega_j(k)$}
 As is clear from Fig.\,\ref{F:3}, for temperatures below $\sim 200\,$K, the main contribution to  Eqs.\,\eqref{S-T} comes from the lowest ferromagnetic mode (da1). As noted in \cite{Cherepanov1993}, the simplified equations yield an accuracy of approximately 5\%, which is sufficient for our purposes, and have the form:
 \begin{subequations}\label{Freq}
 \begin{align} \label{FreqA} \nonumber 
 \omega_{\rm da1}  (\bm k )\approx \,\,&  5 |J_{\rm ad}| \Big \{ \Big [1+ 40 \Big (1 - \frac{J_{\rm dd}}{2J_{\rm ad}} - 2 \frac{J_{\rm aa}}{J_{\rm ad}} \Big )q^2  \\ \nonumber 
  - \big( & 28    -     2  f_4 (\bm n) \big)  q^4   \Big ]^{1/2}    -1 -2 \big ( \frac{J_{\rm dd}}{J_{\rm ad}} -4 \frac{J_{\rm aa}}{J_{\rm ad}}  \big ) q^2 \Big \}\,, \\
\omega_{ ad1} (\bm k )\approx \,\,& \omega_{\rm da1} (\bm k ) - 5 J_{\rm ad}\,,\quad q= (a k)/8\,,  \\ \nonumber 
    f_4 (\bm n)= \,\,& 3 \big ( n_x^2 n_y^2 + n_x^2 n_z^2 +n_y^2 n_z^2 \big )\ .
 \end{align} 

 By disregarding small anisotropic contributions ($0 \leq f_4 \leq 1$) and using the numerical values \eqref{int} of the exchange integrals, one finds 
 \begin{align}\begin{split}\label{Frec1A}
 \omega_{\rm da1}  (\bm k ) \approx  &\,\, 400 \Big \{ \sqrt{ 1+ 25.4\, q^2 -27 q^4   } \\
 & -1 +0.07   q^2 \Big \}\,\mbox{K}\,, \\ 
  \omega_{\rm ad1}  (\bm k ) \approx & \,\Big [ 400 +  \omega_{\rm da1}  (\bm k )\Big ]\,\mbox{K}
\end{split}\end{align}
Equations\, \eqref{Freq} shows that for small $q$
\begin{align}\begin{split}\label{frec1}
\omega(k)=&\,\,\omega_{\rm ex} (ak)^2\,,  \\\omega_{\rm ex} = \,&\frac 5{16}(8 J_{\rm aa}+ 3 J_{\rm dd}- 5 J_{\rm ad})\approx 80 \,\mbox{K} \,, 
\end{split}\end{align} 
 \end{subequations}
in agreement with the well known result\,\cite{Douglass1960}. We see that the contribution of the exchange interaction to the frequency of the ferromagnetic branch \eqref{frec1} vanishes in the long spin-wave limit $k\to 0$. In this limit, one has to account for the contribution of the magnetic dipole-dipole interaction, which is independent of the magnitude of $k$, being dependent only on the direction of vector $\bm k$ via  $\Theta_{\bm k}$:
\begin{equation}\label{om}
\omega(\bm k)= g (H_0 -  4 \pi M N_z + 2 \pi M \sin^2 \Theta_{\bm k}) + \omega_{\rm ex} (ak)^2\,,
\end{equation}
see, e.g. Eq.(3.1.10) in \cite{Lvov2012}. Here,  $g$ is the gyromagnetic factor, $H_0$ is the value of the external magnetic field,   $\Theta_{\bm k}$ is the angle between $\bm k$ and $\bm M$, and  $N_z$, as before,  is the demagnetization factor ($N_z=1/3$ for spherical samples).  

Equation\, \eqref{om} allows one to estimate the crossover wavenumber $k_*$  at which the contributions of the dipole-dipole and exchange interactions to the frequency become equal: $\omega_{\rm ex}(ak_*)^2= 4\pi g M$. In YIG at room temperature, this gives $ ak_* \approx 0.07$\cite{Lvov2012}.  This value is significantly smaller than the size of the Brillouin zone (with $ak \sim 5$, see Fig. \ref{F:3}). Consequently, the contribution of very long spin waves, where dipole-dipole interactions become important, can be safely ignored. This enables us to omit them from the integrals in \eqref{not} for ${\cal N}_\pm$, which determine, according to Eqs.\eqref{S-T}, the spin-wave suppression of magnetization in our approach.

Returning back to \eqref{frec1}, notice that  for $q> q_*$, above some relatively small transition value $q_*= 1/ \sqrt {25.4}\approx 0.2$, $\omega(k)$ becomes almost linear in $k$
\begin{align} \label{frec2}   
\omega _{\rm da1}\approx (-\omega_{\rm neg} + \omega_{\rm lin}\, ak)\,,  
 \end{align}

 as seen in Fig.\,\ref{F:3}. In \eqref{frec2}, $\omega_{\rm neg} \approx 110\,$K  and   $\omega_{\rm lin}\approx 170\,$K.  The nearly linear part of the dispersion laws $\omega _{\rm da1}$  and $\omega _{\rm ad1}$  originates from the fact that YIG is close to the antiferromagnetic arrangement~\footnote{In YIG, the ratio of the numbers of oppositely oriented spins is $12/8= 3/2$. Their difference $12-8 = 4$ is small with respect to half of their sum $=10$ ($4/10=0.2$). In this sense, the YIG magnetic structure is relatively close to AFM, where the dispersion laws exhibit a similar form~\cite{Lvov2012}.}

For the rest of the magnon frequencies, it would be sufficient to accept  the parabolic approximation 
\begin{equation}\label{approx-omega}
\omega_j(k) = \omega_{0,j} + \omega_{{\rm ex}, j}(ak)^2\ .
\end{equation}
The values of $ \omega_{0,j}$ and  $\omega_{\rm ex}$ in the left part of the Tab.\,\ref{t:2} are estimated from the corresponding curves of of $ \omega_{j}(\bm k)$, shown in Fig.\,\ref{F:3}.  To estimate  $ \omega_{0,j}$ and  $\omega_{{\rm ex},j }$ shown in the right part of Tab.\,\ref{t:2}, we used  $ \omega_{j}(\bm k)$  presented in Refs.\,\cite{Princep2017a,Princep2017j}.

\section{\label{A3} Measuring Procedure and Apparatus}

The magnetization measurements for this work were carried out using a Physical Property Measurement System (PPMS, \emph{Quantum Design}), which allows temperature control from 400\,K down to 1.8\,K and enables the application of external magnetic fields of up to 90\,kOe, generated by a pair of superconducting coils in Helmholtz configuration. The measurements were performed using the vibrating sample magnetometer (VSM) option integrated into the PPMS as an instrument insert.

A commercially obtained YIG sphere (\emph{Microspheres,Inc}) with a diameter of 1\,mm was fixed to the sample holder using nonmagnetic adhesive, with the easy axis—the $\langle111\rangle$ crystallographic direction—aligned parallel to the applied external magnetic field. During the measurements, a constant external field of 10\,kOe was applied, which is sufficient to ensure magnetic saturation of the YIG sphere across the entire available temperature range.

The sample was slowly cooled down and warmed up at a rate of 0.25\,K/min to ensure equilibrium at every temperature. The magnetization data presented represent the average of measurements taken during both the cooling and warming cycles.

To accurately determine the sample volume for calculating the magnetization from the measured magnetic moment, the mass of the YIG sphere was determined using a high-precision scale and converted into volume using the literature value for the density of YIG.

\section{\label{A6} Numerical Methods}
All numerical calculations presented in this work were carried out using \textit{Wolfram Mathematica} version 13.0~\cite{Mathematica}.

For the fit shown in Fig.\,\ref{F:2}, represented by the violet curve \textcircled{4}, we used the experimental data from Ref.~\cite{Hansen1974} [orange dots \textcircled{2}]. The high-temperature fit was based on the phenomenological expression $\overline{S}^{z} = A  (T_{_{\rm C}} - T)^\beta$, where $A$ and $\beta$  are fit parameters. The fitting was performed using the \texttt{NonlinearModelFit} function in \textit{Mathematica}. To ensure a reliable result, it was necessary to carefully select the temperature range for the fit. Data in the interval from 350\,K to 535\,K was chosen. The upper bound had to remain below the Curie temperature $  T_{_{\rm C}} = 560\,\text{K} $, as data too close to $T_{_{\rm C}} $ is affected by an experimental artifact—residual magnetization above $T_{_{\rm C}} $ caused by measurements taken in a relatively strong external magnetic field of 10\,kOe. Conversely, the lower bound was set near room temperature, as the fitting model does not account for the saturation of magnetization at lower temperatures; thus, including such data would distort the fit parameters. It is worth noting that the fit results remained consistent under variations of the chosen interval limits by $ \pm 10\,\text{K}$.

The curves $[-\overline{S}^{z}_\mathrm{a}(T)/\overline{S}^{z}_\mathrm{a}(0)]$ \textcircled{9} and $\overline{S}^{z}_\mathrm{d}(T)/\overline{S}^{z}_\mathrm{d}(0)$ \textcircled{10} in Fig.\,\ref{F:4}, as well as their difference $\overline{S}^{z}(T)/\overline{S}^{z}(0)$ \textcircled{5} in Figs.\,\ref{F:2} and \ref{F:4}, were obtained by numerically solving the system of Eqs.\,\eqref{4} using the parameters specified in Eq.\,\eqref{cal-J}. The numerical solution was carried out using the \texttt{FindRoot} function in \textit{Mathematica}.

For the curves \textcircled{7} in Fig.\,\ref{F:2} and the additional curves \textcircled{11} and \textcircled{12} in Fig.\,\ref{F:5}, Eq.\,\eqref{S-Te} was evaluated. This required calculating the total number of magnons within the first Brillouin zone as given by Eq.\,\eqref{S-Tf}, which involves an integration over the wavevector $\bm{k}$. In this work, we approximate the first Brillouin zone by a sphere of radius equal to the Debye wavevector $k_{\rm D}$, which for a BCC lattice is given by $k_{\rm D} = \sqrt[3]{12\pi^2/a^3}$, a simplification that provides a good approximation for our purposes. Accordingly, the integral defined in Eq.\,\eqref{not} over $n_j(\bm{k})$, Eq.\,\eqref{BE}, was performed in spherical coordinates, using the magnon dispersion relations $\omega_j(\bm{k})$ provided in Appendix~\ref{E}. The integration range was set from $k_{\rm min} = 0$ to $k_{\rm max} = k_{\rm D}$. The numerical evaluation was carried out using the \texttt{NIntegrate} function in \textit{Mathematica}.

The main result of this work, curve \textcircled{8} (red solid line) in Fig.\,\ref{F:2}, was obtained by solving Eq.\,\eqref{26}. The magnon occupation numbers ${\cal N}_{+}$ and ${\cal N}_{-}$ were calculated as described in the previous paragraph, but explicitly using temperature-dependent dispersion relations given by Eq.\,\eqref{T-dep1}, with $ \delta = 0.428$. The equation was numerically solved for $\overline{S}^{z}$ using the \texttt{FindRoot} function in \textit{Mathematica}.


 %

\end{document}